%% file: TIT_publish.tex
\documentclass[journal]{IEEEtran}

\usepackage{amsmath,amssymb,float,arydshln,color}
\usepackage{psfrag,setspace,wrapfig,subfigure}
\usepackage[latin1]{inputenc}
\usepackage{dsfont}
\usepackage{epsfig}
\usepackage{epstopdf}
\usepackage{graphicx}
\usepackage{hyperref}
\usepackage{cite}
\usepackage{url}
\allowdisplaybreaks

\input{macros}

\begin{document}
\def\helvetica{phvr7t.tfm}
\def\helveticaoblique{phvro7t.tfm}
\def\helveticabold{phvb7t.tfm}
\def\helveticaboldoblique{phvbo7t.tfm}
\font\sfb=\helveticabold
=\helveticaboldoblique
\title{Information Exchange and Learning Dynamics over Weakly-Connected Adaptive Networks}

\author{Bicheng~Ying,~\IEEEmembership{Student Member,~IEEE},~and Ali~H.~Sayed,~\IEEEmembership{Fellow,~IEEE}
\thanks{This work was supported in part by NSF grants CIF-1524250 and ECCS-1407712. A short version of this word appeared in the conference publication \cite{ying2015learning}.

The authors are with Department of Electrical Engineering, University of California, Los Angeles, CA 90095. Emails: \{ybc,sayed\}@ucla.edu.}}
\maketitle
\begin{abstract}
The paper examines the learning mechanism of adaptive agents over weakly-connected graphs and reveals an interesting behavior on how information flows through such topologies. The results clarify how asymmetries in the exchange of data can mask local information at certain agents and make them totally dependent on other agents. A leader-follower relationship develops with the performance of some agents being fully determined by the performance of other agents that are outside their domain of influence. This scenario can arise, for example, due to intruder attacks by malicious agents or as the result of failures by some critical links. The findings in this work help explain why strong-connectivity of the network topology, adaptation of the combination weights, and clustering of agents are important ingredients to equalize the learning abilities of all agents against such disturbances. The results also clarify how weak-connectivity can be helpful in reducing the effect of outlier data on learning performance.
\end{abstract}

\begin{keywords}
Weakly-connected graphs, distributed strategies, information exchange, Pareto optimality, leader-follower relationship, limit points, outlier data, malicious agents. 
\end{keywords}

\section{Introduction}
This work examines the distributed solution of inference problems by a collection of networked agents. An individual convex cost function  $J_k(w):\real^{M}\rightarrow \real$ is associated with each agent $k=1,2,\ldots,N$, and the objective is for the agents to cooperate locally in order to determine the global minimizer of the aggregate cost. Several useful techniques have been developed in the literature for this purpose, including the use of consensus strategies \cite{kar2012distributed,Nedic09,dimakis10,boyd2006randomized,tsianos2012consensus,kar2013distributed,stankovic2011decentralized} and diffusion strategies \cite{Sayed14,Sayed14b,sayedSPM}. In most prior studies (see, e.g., \cite{Nedic09,dimakis10,kar2012distributed,Sayed14,Sayed14b,sayedSPM,chouvardas2011adaptive, braca2008running,dini2012cooperative,takahashi2010link,barbarossa2013distributed,bertrand2013seeing,namvar2013distributed,tsianos2012distributed,ram2010distributed,predd2009collaborative,xiao2004fast,tsitsiklis1986distributed,stankovic2011decentralized,tsianos2012consensus,kar2013distributed,boyd2006randomized,saligrama2006distributed,bianchi2012performance}), it has been generally assumed that the network topology is strongly-connected, which means that a path can be found linking any pair of agents and, moreover, at least one agent has a self-loop. In this case, and under some mild technical conditions \cite{Sayed14}, it is known that all agents are able to approach the global minimizer within $O(\mu_{\max})$, where $\mu_{\max}$ denotes the largest step-size parameter used by the adaptive agents. This means that strongly-connected agents are able to learn well, with all agents attaining a similar performance level despite possible variations in the signal-to-noise ratios across the agents.  

The main theme of the current article is to examine how the learning behavior of the agents is affected when the network topology is not necessarily strongly-connected. Specifically, we shall examine the effect of weak-connectivity on information flow, where a weakly-connected network is one that consists of multiple clustered agents with at least one cluster always feeding information forward but never receiving information back from any of the other clusters. Such graph settings arise in important situations. For example, they can arise as the result of intruder attacks by 
malicious agents who keep feeding their neighbors with inaccurate information but never use the information fed back to them by the other agents. This behavior results in a unilateral and asymmetric information exchange scenario, which can be modeled by weakly-connected networks. A second example arises in the context  of interactions over social networks where weak connectivity can be used to model the behavior of stubborn agents that insist on their opinions regardless of the feedback by others, or to model the ability of organizations to control the flow of media information \cite{acemoglu2011opinion,yildiz2011discrete}. A third example arises in the context of information dissemination over social platforms such as Twitter and Weibo, where celebrities may have thousands of followers while these special users may be following only a small subset of trusted acquaintances.  All these examples lead to heavily biased asymmetric information exchange scenarios, which can be modeled reasonably well by weakly-connected networks.

Motivated by these considerations, we examine the learning mechanism of adaptive agents over weakly-connected graphs in some detail and reveal an interesting behavior. We will find that a leader-follower relationship develops among the agents, with the performance of some agents being fully determined/controlled  by the performance of other agents that are outside their domain of influence. This scenario does not only arise as the result of intruder attacks or highly asymmetric information exchanges, as explained before, but can also be the  the result of failures by some critical links that render the network topology weakly-connected.  Among other contributions, the findings established in this article will help explain why strong-connectivity of the network topology \cite{Sayed14,Sayed14b} adaptation of the combination weights  \cite{Sayed14,sayedSPM} and clustering of agents\cite{zhao15distributed,CRS14} are important ingredients to safeguard against such pitfalls. The conclusion will also highlight one useful scenario where weak-connectivity is beneficial, namely, in reducing the effect of outlier data on network performance. 

 Some useful related works on leader-follower relationships appear in \cite{khan2010higher,abaid2012leader,PhysRevE.86.036105,liu2008controllability}. References \cite{abaid2012leader,PhysRevE.86.036105} are primarily motivated by the
behavior of fish schools and the switching between leader and follower agents, while \cite{liu2008controllability} is concerned with controllability issues in the context of 
a switching topology. The work \cite{khan2010higher} also deals with weakly-connected networks but differs critically from our formulation in that 
it considers a {\em static} scenario with anchor sensors and initial {\em fixed} state values. The purpose is to show that an average consensus
iteration will make the other sensors converge to a combination of the anchor states. While this is a useful conclusion, it is not surprising  that it holds 
in this scenario due to the static nature of the problem formulation. The key difference in relation to the work proposed in the current manuscript  is that we will be dealing
with a {\em dynamic} network scenario where data are constantly streaming into the agents. There are no anchor agents or fixed state values. Instead, the data are continuously 
changing. Besides, the probability distribution of the data is unknown and the agents need to rely on approximate gradient information, as opposed to actual
state values, to carry out their learning. Under such circumstances, gradient noise will be constantly present in the network and will seep into the operation of the distributed strategies. The
topology couples the agents together and these gradient noise sources end up diffusing through the network. In this case, it is much less trivial to identify the limit points
of the various sub-networks and agents. This is because the presence of persistent gradient noise now forces agents to fluctuate around 
limits points. The challenge is to show that the dynamics over the network is stable enough to keep these noises under check and to lead to a steady-state
scenario with well-defined (mean-square-error) limit points. The analysis in the paper will show that this is indeed the case and will derive two main results 
(Theorems \ref{theorem.1} and \ref{theorem.3}), which characterize analytically and in closed-form the stability and performance of weakly-connected graphs under dynamic continuous learning.


{\em Notation}: We use lowercase letters to denote vectors, uppercase
letters for matrices, plain letters for deterministic
variables, and boldface letters for random variables. We also
use $(\cdot)^{\sf T}$  to denote transposition, $(\cdot)^{-1}$ for matrix inversion,
$\mbox{\sf Tr}(\cdot)$ for the trace of a matrix, $\lambda(\cdot)$ for the eigenvalues of
a matrix, $\|\cdot\|$ for the 2-norm of a matrix or the Euclidean
norm of a vector, and $\rho(\cdot)$ for the spectral radius of a matrix.
Besides, we use $ A \geq  B$ to denote that $A - B$ is positive
semi-definite, and $p\succ 0$ to denote that all entries of vector $p$ are positive. 

\section{Strongly-Connected Networks}
In preparation for the derivation of the main results in this work, we first review the setting of strongly-connected networks following \cite{Sayed14,Sayed14b}.

\begin{figure}[h]
\epsfxsize 9cm \epsfclipon
\begin{center}
\leavevmode \epsffile{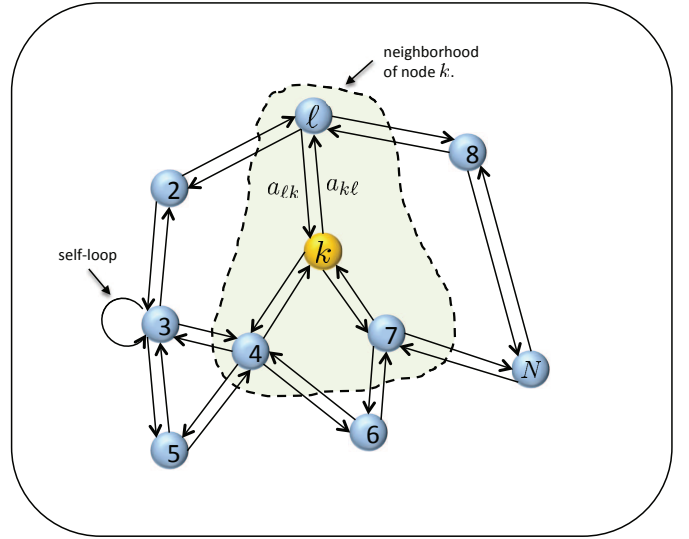} \caption{{\small Agents
that are linked by edges can share information. The neighborhood of agent
$k$ is marked by the broken line and consists of the set
${\cal N}_k=\{4,7,\ell,k\}$.}}\label{fig-A.label}
\end{center}
\end{figure}

Thus, consider a network consisting of $N$ separate agents connected by a topology. We assign a pair of nonnegative weights, $\{a_{k\ell},a_{\ell k}\}$, to the edge connecting any two agents $k$ and $\ell$. The scalar $a_{\ell k}$ is used by agent $k$ to scale the data it receives from agent $\ell$ and similarly for $a_{k\ell}$. The weights $\{a_{k\ell},a_{\ell k}\}$ can be different, and one or both weights can also be zero. The network is said to be {\it connected} if paths with nonzero scaling weights can be found linking any two distinct agents in both directions. The network is said to be {\em strongly--connected} if it is connected with at least one self-loop, meaning that $a_{kk}>0$ for some agent $k$. In this way, information can flow in both directions between any two distinct agents and, moreover, some vertices possess self-loops with positive weights \cite{Sayed14}. Figure~\ref{fig-A.label} shows one example of a strongly--connected network. For emphasis in this figure, each edge between two neighboring agents is represented by two directed arrows. The neighborhood of any agent $k$ is denoted by ${\cal N}_k$ and it consists of all agents that are connected to $k$ by edges; we assume by default that this set includes agent $k$ regardless of whether agent $k$ has a self-loop or not.

For each agent $k$, the nonnegative weights it employs to scale data from its neighbors will be convex combination coefficients satisfying:
\be
a_{\ell k}\geq 0,\;\;\;\sum_{\ell\in{\cal N}_k} a_{\ell k}=1,\;\;\;\;a_{\ell k}=0\;\mbox{\rm if}\;\ell\notin{\cal N}_k
\label{label.eq1}\ee
Assume we collect the  coefficients $\{a_{\ell k}\}$ into an $N\times N$ matrix $A=[a_{\ell k}]$. Then, condition (\ref{label.eq1}) implies that $A$ satisfies
\be
A\tran \one =\one
\ee
so that $A$ is a left-stochastic matrix. Additionally, the strong connectivity of the network implies that $A$ is a primitive matrix \cite{Sayed14}.  One important property of left-stochastic primitive matrices follows from the Perron-Frobenius Theorem \cite{Horn03,Pillai05}; it asserts that the matrix $A$ will have a single eigenvalue at one, with all other eigenvalues lying strictly inside the unit circle. Moreover, if we let $p$ denote the right-eigenvector of $A$ corresponding to its single eigenvalue at one, and normalize its entries to add up to one, then all entries of $p$ will be strictly positive, meaning that $p$ satisfies:
\be
Ap=p,\;\;\;\;\one\tran p=1,\;\;\;\;p\succ 0\label{label.eq3}
\ee
We refer to $p$ as the Perron eigenvector of $A$.

\subsection{Aggregate Cost Function}
We associate with each agent, $k$, a twice-differentiable and convex cost function, denoted by $J_k(w)\in\real$, with independent variable $w\in\real^{M}$. We assume at least one of these costs is strongly-convex. The agents run a collaborative distributed strategy of the following adapt-then-combine (ATC) diffusion form:
\bq
          \bm{\psi}_{k,i}  &=&   \displaystyle \w_{k,i-1} - \mu_k
          \,\widehat{\nabla_{w\tran} J}_k(
                    \w_{k,i-1})\label{label.eq4a}\\
                    \w_{k,i} &=&   \displaystyle \sum_{\ell \in \mathcal{N}_k} a_{\ell k}\; \bm{\psi}_{\ell,i}\label{label.eq4b}
\eq
where $\mu_k$ is a positive step-size at agent $k$, the vectors $\{\bm{\psi}_{k,i},\w_{k,i}\}$ denote iterates at agent $k$ at time $i$, and
$\widehat{\nabla_{w} J}_k(\w_{k,i-1})$ represents an approximation for the true gradient vector of $J_k(w)$; an approximation is needed because, over adaptive networks, the true cost functions $J_k(w)$ are generally not known in advance and their gradients need to be estimated continually from streaming data\cite{Sayed14,Sayed14b}. The difference between the true gradient vector and its approximation is called gradient noise --- see (\ref{label.eqRR}) further ahead.  Other strategies of the consensus or diffusion type are possible (e.g., \cite{Sayed14,Sayed14b,Nedic09,dimakis10}), but it is sufficient to illustrate the results using (\ref{label.eq4a})--(\ref{label.eq4b}) \cite{Sayed14,tu2012diffusion}.
We represent the step-sizes as scaled multiples of the same factor $\mu_{\max}$, namely, \be \mu_k\define \tau_k\, \mu_{\max},\;\;\;\;k=1,2,\ldots,N\label{label.eq5}\ee
where $0<\tau_k\leq 1$. We also introduce the vector:
\be q\define \mbox{\rm diag}\{\mu_1,\mu_2,\ldots,\mu_N\}\cdot p\label{label.eq6} \ee
whose individual entries, $\{q_k,\,k=1,2,\ldots,N\}$, are obviously positive.
Using the $\{q_k\}$, we define the strongly-convex weighted aggregate cost: \be
J^{\rm glob,\star}(w)\;\define\;\sum_{k=1}^N q_k J_k(w)
\label{label.eq7}\ee

\subsection{Mean-Square-Error Performance}
We denote the unique minimizer of (\ref{label.eq7}) by $w^{\star}$ and measure the error at each agent relative to this limit point by means of the vectors:
\be
\widetilde{\w}_{k,i}\define w^{\star}-\w_{k,i},\;\;\;\;k=1,2,\ldots,N\label{label.eq8}
\ee
It was shown in \cite{Sayed14,Sayed14b,Chen13,chen2015learning1,chen2015learning2} that $w^{\star}$ serves as a Pareto optimal solution for the network. Specifically, under some reasonable technical conditions on the cost functions and gradient noise process, it holds that (see Theorem 9.1 of \cite{Sayed14}):
\bq
\limsup_{i\rightarrow\infty}\,\Ex\|\widetilde{\w}_{k,i}\|^2&=&
O(\mu_{\max})\label{label.eq9}
\eq
That is, all agents approach in the mean-square-error sense the  limiting point $w^{\star}$ to within $O(\mu_{\max})$ for sufficiently small step-sizes. We let $\mbox{\rm MSD}_k$ denote the size of the mean-square deviation, $\Ex\|\widetilde{\w}_{k,i}\|^2$, in steady-state to {\it first-order} in $\mu_{\max}$. We also let
$\mbox{\rm MSD}_{\rm av}$ denote the average mean-square deviation across all $N$ agents.
It was further shown in \cite{Sayed14,Sayed14b,chen2015learning2} that these measures are given by (see Lemma 11.3 of \cite{Sayed14}):
\be \mbox{\rm MSD}_{k}=\mbox{\rm MSD}_{\rm av}=\frac{1}{2}\mbox{\rm Tr}\left[
\left(\sum_{k=1}^N q_k H_k\right)^{-1}\left(\sum_{k=1}^N q_k^2
G_k\right)
\right]\label{label.eq10}\ee
where the $M\times M$ matrix quantities $\{H_k,G_k\}$ correspond to the Hessian matrix of the cost function and to the covariance matrix of the gradient noise process, respectively, at agent $k$:
\bq
H_k&\define&\nabla_w^2\;J_k(w^{\star})\label{label.eq11a}\\
G_k&\define&\lim_{i\rightarrow\infty}
\Ex\left[\,\s_{k,i}(w^{\star})\s_{k,i}\tran(w^{\star})\,|\,\bm{\cal F}_{i-1}\,\right]
\label{label.eq11b}\eq
with the symbol $\s_{k,i}$ used to denote the gradient noise process:
\be
\s_{k,i}(\w_{k,i-1})\define \widehat{\nabla_{w\tran}J}_k(\w_{k,i-1})\;-\;
\nabla_{w\tran} J_k(\w_{i-1})
\label{label.eqRR}\ee
and where $\bm{\cal F}_{i-1}$ denotes the filtration corresponding to all past iterates across all agents:
\be \bm{\cal F}_{i-1}\;=\;\mbox{\rm filtration by $\{\w_{\ell,j},\;j\leq i-1,\;\ell=1,2,\ldots,N\}$}
\ee

\section{Weakly-Connected Networks}
\begin{figure}[h]
\epsfxsize8.5cm \epsfclipon
\begin{center}
\leavevmode \epsffile{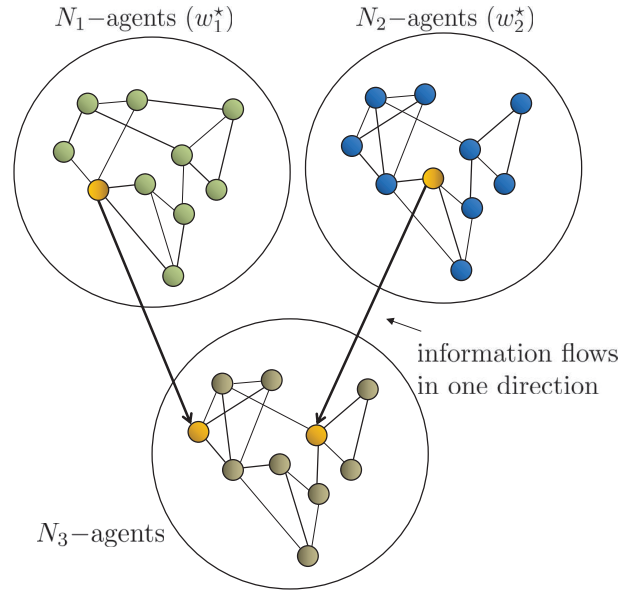} \caption{{\small Illustration of a weakly-connected network consisting of three sub-networks. }}\label{fig-B.label}
\end{center}
\end{figure}

We now examine the same learning mechanism over weakly-connected networks. The objective is to verify whether some agents can end up having a dominating effect on the performance of other agents. Broadly, a weakly-connected network consists of a collection of sub-networks with constraints on how information flows among them. Figure~\ref{fig-B.label} illustrates one particular situation consisting of three sub-networks, with the number of their agents denoted by $N_1$, $N_2$, and $N_3$, respectively. These numbers can be equal to each other or they can be different. Although Figure~\ref{fig-B.label} is limited to three sub-networks, the results and analysis in the sequel are general and apply to any number of sub-networks. The figure is used for illustration purposes only.   There can be many more $R-$type (receiving) sub-networks, as well as many more
$S-$type (sending) sub-networks.

In the figure, each of the two sub-networks on top is strongly-connected and does not receive information from any other sub-network (self-loops are not indicated in the figure). Each of these sub-networks has at least one strongly-convex cost and its own combination policy, denoted by $\{A_1\in\real^{N_1\times N_1},\,A_2\in\real^{N_2\times N_2}\}$, and its own Perron vector, denoted by $\{p_{1},p_2\}$. Therefore, if each of these sub-networks were to run the diffusion strategy (\ref{label.eq4a})--(\ref{label.eq4b}), then each one of them will independently converge in the mean-square-error sense towards its own Pareto solution, denoted by $\{w_1^{\star},w_2^{\star}\}$. The same figure shows a third sub-network in the bottom, and which appears at the {\em receiving} end relative to the other sub-networks. The figure shows two arrows emanating in one direction from the top sub-networks towards the bottom sub-network. Therefore, this third sub-network is influenced by the behavior of the top sub-networks, while it does not feed any information back to them. We would like to examine how the limiting behavior of this third sub-network is influenced by the two top sub-networks and whether it can still exhibit independent behavior. This is an important question with critical implications for inference over networks. We will discover that in situations like these, where one or more sub-networks are at the receiving end of other sub-networks, a leader-follower relationship develops with the limiting behavior of the receiving sub-networks being completely dictated by the other sub-networks regardless of their own local information (or opinions).


\subsection{Network Model}
Thus, consider a network consisting of a collection of $S$ stand-alone strongly-connected sub-networks. Each of these sub-networks does not receive information from any other sub-network and they can therefore run their diffusion strategy independently of the other sub-networks. We further assume that the network contains a second collection of $R$ sub-networks where some agents in these sub-networks can receive information from agents in the first collection. The letters $S$ and $R$ are chosen to designate ``send'' and ``receive'' effects: sub-networks in group $S$ have agents that send information to sub-networks in the receiving group $R$. Whenever necessary, we will use the small letters $s$ and $r$ as subscripts to refer to sub-networks or quantities from group $S$ and to sub-networks or quantities  from group $R$. If we refer to Figure~\ref{fig-B.label}, then $S=2$ and $R=1$. The total number of agents in the network is still denoted by $N$ and it is equal to the sum of the number of agents across all sub-networks.\footnote{We remark that our definition of weakly-connected networks is more strict than the terminology used  in graph theory. There, a directed graph is called weakly-connected if replacing all of its directed edges with undirected edges produces
a connected (undirected) graph \cite{bondy1976graph}. This definition would also include strongly-connected networks as special cases. Our definition is meant to focus  on networks that are truly 
weakly-connected in that they induce an asymmetric flow of information among some of its components.}

We collect all weighting coefficients $\{a_{\ell k}\}$ from across all edges into a large $N\times N$ combination matrix $A=[a_{\ell k}]$.
Without loss of generality, we assume the agents are numbered with the agents from the union of all $S$ strongly-connected sub-networks coming first, followed by the agents from the remaining $R$ sub-networks. 

A useful matrix decomposition result proven in \cite[Ch. 8]{Meyer00} shows that the combination matrix of every such weakly-connected network can be transformed, via a symmetric permutation transformation of the form $P\tran AP$, to an upper block-triangular structure of the following form (in other words, the assumed structure (\ref{label.eq20}) is general enough to represent any weakly-connected graph):

\begin{small}
	\begin{eqnarray}
	\nonumber
	\begin{array}{cc}
	\overbrace{\rule{30mm}{0mm}}^{\mathrm{Subnetworks:} 1,2,\ldots, S} &\; \overbrace{\rule{50mm}{0mm}}^{\mathrm{Subnetworks:} S+1, S+2, \ldots,S+R}
	\end{array}
	\\ \nonumber
	\left[
	\begin{array}{cccc|cccc}
	A_{1} 	&	0 	     	&\hdots	& 0 		&	A_{1,S+1} 		& A_{1,S+2}		 &\hdots		 &A_{1,S+R}	\\
	0          	& A_{2}   		&\hdots	& 0 		& 	A_{2,S+1} 		& A_{2,S+2}	 	 &\hdots		 &A_{2,S+R}	\\
	\vdots 	& \vdots		&\ddots	&\vdots	& 	\vdots 			& \vdots			 &\ddots		 &\vdots	 \\
	0          	& 	0  		&\hdots	& A_{S} 	& 	A_{S,S+1} 		& A_{S,S+2}	 	 &\hdots		 &A_{S,S+R}	\\
	\hline
	0          	& 	0  		&\hdots	& 0 		& 	A_{S+1} 			& A_{S+1,S+2}	 	 &\hdots		 &A_{S+1,S+R}	\\
	0          	& 	0  		&\hdots	& 0 		& 	0		 		& A_{S+2}	 	 	 &\hdots		 &A_{S+2,S+R}	\\
	\vdots 	& \vdots		&\ddots	&\vdots	& 	\vdots 			& \vdots			 &\ddots		 &\vdots	 \\
	0          	& 	0  		&\hdots	& 0 		& 	0				& 0			 	 &\hdots		 &A_{S+R}	 \\
	\end{array}
	\right]
	\end{eqnarray}
\end{small}
\be
\ \label{label.eq20}
\ee
\noindent In the above expression, the quantities $\{A_{1},  \ldots,  A_{S}\}$ are the left-stochastic primitive matrices corresponding to the $S$ strongly-connected sub-networks. We sometimes denote a generic $A_s$ from this set by $A_{s,s}$ to mean that it is the combination matrix from sub-network $s$ to itself. We denote the size of each
matrix $A_s$ by $N_s$.

 Likewise, the matrices
$\{A_{S+1},  \ldots,  A_{S+R}\}$ in the lower right-most block contain the internal combination coefficients for the $R$ collection of sub-networks. For example, $A_{S+1}$ contains the coefficients that appear on the edges within sub-network $S+1$; this matrix is {\em not} left-stochastic because it does not contain all the combination coefficients that are used by the agents within sub-network $S+1$. For example, $A_{S+1}$ does not contain any of the coefficients on the edges arriving to sub-network $S+1$ from any of the first $S$ sub-networks. The zero entries in the lower-left block corner refer to the fact that none of the $S$ sub-networks receive information from the $R$ sub-networks. The entries in the right-most upper block contain the combination weights from the edges that emanate from the $S$ sub-networks towards the $R$ sub-networks.

\begin{figure}[h]
\epsfxsize 5.2cm \epsfclipon
\begin{center}
\leavevmode \epsffile{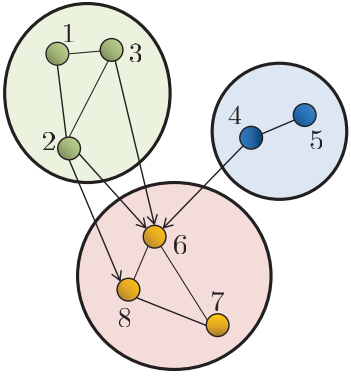} \caption{{\small A weakly connected network consisting of three sub-networks and the corresponding combination policy (\ref{label.eq11}). }}\label{fig-C.label}
\end{center}
\end{figure}

For example, for the network shown in Figure~\ref{fig-C.label}, one possibility for $A$ is:

\begin{small}
\begin{equation}
A=\left[
	\begin{array}{ccccc|ccc}
		0.2		&	0.2 	     	&0.8		& 0 		&	0 		& 0		&0		&0	 \\
		0 .5   	&      0.4   		&0.1		& 0 		& 	0		& 0.2 	&0		 &0.4	 \\
		0.3 		& 	0.4		&0.1		&0		& 	0 		& 0.1		&0		&0	\\
		0          	& 	0  		&0		& 0.4 	& 	0.3 		& 0.3 	&0		&0	\\
		0          	& 	0  		&0		& 0.6 	& 	0.7 		& 0	 	&0		&0	\\
		\hline
		0          	& 	0  		&0		& 0 		& 	0		& 0.2	 	&0.3		&0.2	 \\
		0		&      0		&0		& 0		& 	0		& 0.1		&0.5		&0.3	 \\
		0          	& 	0  		&0		& 0 		& 	0		& 0.1		&0.2		&0.1	 \\
	\end{array}
\right]
\label{label.eq11}\end{equation}
\end{small}

\section{Steady-State Dynamics}
We are interested in examining the steady-state behavior of each agent in the network. To do so, we will need to examine first the limit of $A^n$ as $n\rightarrow\infty$, for matrices $A$ that have the structure (\ref{label.eq20}).

\subsection{Limiting Power of Combination Matrix}
We denote the block structure of $A$ from  (\ref{label.eq20}) by
\be
A\define \ba{ccc}T_{SS}&\vline&T_{SR}\\\hline 0&\vline &T_{RR}\ea\label{eq.thisx}
\ee
where, for example, $T_{SS}$ is block diagonal and consists of the left-stochastic and primitive entries $\{A_1,A_2,\ldots,A_s\}$, while $T_{RR}$ is block upper triangular. The block $T_{SR}$ represents the influence of the $S$ sub-networks on the $R$ sub-networks.

\begin{lemma}[{\sc Limiting power of $A$}] Let  the Perron eigenvectors of the $S$ strongly-connected sub-networks be denoted by $\{p_{s},\,s=1,2,\ldots,S\}$. It holds that:
\begin{equation}
   A_{\infty} \stackrel {\triangle}{=}
    \lim_{n\to\infty} A^n =
    \left[
      \begin{array}{ccc}
        \Theta &\vline& \Theta W \\\hline
        0 &\vline& 0 \\
      \end{array}
    \right]
\label{label.eq30}\end{equation}
where the matrices $\Theta$ and $W$ are defined by
\bq
    W &\define & T_{SR}(I-T_{RR})^{-1} \label{eq:defW}\\
     \Theta &\define &\mbox{\rm blockdiag}\left\{ p_1\one_{N_1}\tran,\ldots,p_S\one_{N_S}\tran\right\}
\eq
\label{kajd713.lemma}\noindent and the notation ``blockdiag'' refers to a block diagonal matrix constructed from its arguments. 
\end{lemma}
\bp We start by establishing that $\rho(A_{r})<1$ for the networks at the receiving end of information, i.e., for $S<r\leq S+R$. Note that we are referring here to the block entries $\{A_r\}$ on the diagonal of $T_{RR}$. Using the fact that the spectral radius of any matrix is upper bounded by any matrix norm,  we have that  $\rho(A_{r}) \leq \|A_{r}\|_1\leq 1$ in terms of the $1-$norm; the property $\|A_r\|\leq 1$ follows from the fact that the columns of $A_r$ are subsets of longer columns whose entries add up to one; refer to the example in (\ref{label.eq11}).

We next show, by contradiction, that strict inequality must hold, i.e., $\rho(A_r)<1$. Thus, assume to the contrary that $\rho(A_{r})=1$. Since $A_r$ has non-negative entries and since sub-network $r$ is connected,
it follows from the Perron-Frobenius Theorem \cite{Horn03,Pillai05} that\footnote{Consider a matrix $A$ consisting of nonnegative entries representing the scaling weights over the edges of a {\em connected} network. Then, in general, $A$ may have multiple eigenvalues that attain $\rho(A)$. The Perron-Frobenius Theorem ascertains that each of these eigenvalues will have multiplicity one, and that only one of them is real, say, denoted by $\lambda$. The theorem further ensures that there exists an eigenvector vector $p$ with positive entries such that $Ap=\lambda p$. If the graph happens to be strongly-connected, instead of only connected, then $\lambda$ is the only eigenvalue that attains $\rho(A)$.} we can find a vector, $p_r$, with positive entries such that $A_r p_r = p_r$. Now note that
\be \one\tran A_{r} \preceq\one \tran\ee
 where the symbol $\preceq$ denotes entry-wise comparisons. Actually, at least one of the entries of the row vector $\one\tran A_r$ must be strictly smaller than one. We then conclude that it must hold:
 \be\one\tran A_r p_r < \one\tran p_r\ee
with strict inequality. For this conclusion to hold, at least one entry of the vector $A_r p_r$  must be smaller than the corresponding entry in $p_r$; this fact contradicts the identity $A_r p_r=p_r$.

We therefore conclude that $\rho(A_r)<1$ and, consequently, $\rho(T_{RR})<1$. Furthermore,  recall that each of the leading blocks $\{A_s,\,s=1,2,\ldots, S\}$ is primitive. Therefore, the powers of each of these matrices tends to \cite{Sayed14,Horn03}:
\be
\lim_{n\rightarrow \infty}\;A_s^n=p_s\one_{N_s}\tran,\;\;\;s=1,2,\ldots,S
\ee
Combining this fact with $\rho(T_{RR})<1$, we conclude that the limit of $A^n$ exists as $n\rightarrow\infty$, and that it has the following generic form:
 \be
	A_{\infty} =
	\left[
	\begin{array}{ccc}
		\Theta &\vline& X \\\hline
		0 & \vline&0
	\end{array}
	\right]
\ee
for some matrix $X$ that we would like to identify.  Using the relationship  $A_{\infty}A = A_{\infty}$ , it holds that
\begin{equation}
	\left[
	\begin{array}{cc}
		\Theta & X \\
		0 & 0
	\end{array}
	\right]
	\left( I -
	\left[
	\begin{array}{cc}
		T_{SS}  & T_{SR} \\
		0 	    & T_{RR}
	\end{array}
	\right]
	\right)
	=
	\left[
	\begin{array}{cc}
		0 & 0 \\
		0 & 0
	\end{array}
	\right]
\end{equation}
from which we conclude that $X = \Theta T_{SR}(I-T_{RR})^{-1}$, as claimed.

\ep

We can provide a useful interpretation for the factor $W$ that appears on the right-hand side of (\ref{label.eq30}). We denote the total number of agents in group $S$ by
\be
	{N}_{gS}\define N_{1}+N_{2}+ \cdots +N_{S}
\ee
and the total number of agents in group $R$ by
 \be
{N}_{gR}\define N_{S+1}+N_{S+2}+\cdots + N_{S+R}
\ee
Then, entries on each column of $W$ add up to one, as can be verified from the following equivalent statements:
\bq
	\one\tran_{N_{gS}} W = \one\tran_{N_{gR}}	 &\stackrel{\tiny (\ref{eq:defW})}{\Longleftrightarrow} &
	\one\tran_{N_{gS}} T_{SR} = \one\tran_{N_{gR}} (I - T_{RR})\nn\\
	&\Longleftrightarrow &\one\tran_{N_{gS}} T_{SR} + \one\tran_{N_{gR}}T_{RR} = \one\tran_{N_{gR}} \nn\\
	&\Longleftrightarrow & \one\tran_N
	\ba{c}
		T_{SR} \\
		T_{RR}
	\ea = \one\tran_{N_{gR}} \label{eq.Wleft}
\eq
where the last step holds because $A$ is left-stochastic.  Moreover, since $T_{RR}$ is a stable matrix, we have
 \bq
 W&=&T_{SR}(I-T_{RR})^{-1}\nn\\
 &=&T_{SR}( I + T_{RR}+T_{RR}^2 + \ldots )\nn\\
 &=&T_{SR} + T_{SR}T_{RR} + T_{SR}T_{RR}^2 +\ldots \label{label.eq40}
 \eq
Recall that $T_{SR}$ represents the combination weights that scale the data emanating from group $S$ and reaching group $R$. In the same token, $T_{RR}$ represents the combination weights that are internal to group $R$ and models how the sub-networks within this group  scale their internal data. As such, the first term in (\ref{label.eq40}) represents the information that is transferred from group $S$ into group $R$, while the second term
 in (\ref{label.eq40}) represents how this information is transformed internally within group $R$ after one step, and similarly for the subsequent terms in (\ref{label.eq40}) involving higher-order powers of $T_{RR}$.

\subsection{Limit Points for Group S of Sub-Networks}
Let $\{w_{s}^{\star},\;s=1,2,\ldots,S\}$ denote the Pareto optimal solutions for the strongly-connected sub-networks in group $S$. In order to characterize the group behavior more fully, we need to introduce a more explicit notation. Thus, consider sub-network $s$ from this group. It has $N_s$ agents and their individual step-sizes will be denoted by $\{\mu_{s,k}\}$, with the first subscript referring to the sub-network and the second subscript referring to the agent. Likewise, the Perron vector of sub-network $s$ will be denoted by $p_s$ with individual entries $\{p_{s,k}\}$. The associated scaled weights are denoted by:
\be
q_{s,k}\define \mu_{s,k}p_{s,k},\;\;\;k=1,2,\ldots,N_s
\label{definqa.s}\ee
According to (\ref{label.eq7}), the Pareto solution, $w_s^{\star}$, that corresponds to sub-network $s$ is the unique solution to the following algebraic equation:
\be
\sum_{k=1}^{N_s}q_{s,k}\nabla_{w\tran} J_{s,k}(w_{s}^{\star})\;=\;0
\ee
where the $\{J_{s,k}(w)\}$ denote the cost functions that are associated with agents $k$ within sub-network $s$. Each agent in sub-network $s$ will converge towards this same $w_{s}^{\star}$ within $O(\mu_{\max})$. This conclusion means that the limit point will be uniform within each sub-network $s$; though the limit points can be distinct across the sub-networks. Collecting the Pareto solutions $\{w_{s}^{\star}\}$ from across the $S$ sub-networks, we find that the limiting points for all agents within group $S$ are described by the following extended vector:
\be
\sw^{\star}\define \ba{c}\one_{N_1}\otimes w_1^{\star}\\
\one_{N_2}\otimes w_2^{\star}\\
\vdots\\
\one_{N_s}\otimes w_S^{\star}
\ea\label{kajdj671381.,1}
\ee
Recall that the total number of agents in group $S$ is $N_{gS}$.

\subsection{Limit Points for Group R of Sub-Networks}
Now consider an arbitrary sub-network $r$ from group $R$. It turns out that, contrary to the uniform behavior observed for group $S$, each agent within the sub-network $r$ will  converge to an individual limit point. This conclusion is established in the sequel (see Theorem~\ref{theorem.1}) and motivated as follows.

First, we recall that the total number of agents across all sub-networks in group $R$ is $N_{gR}$. We denote the limiting value for each agent $k$ in sub-network $r$ by
$w_{r,k}^{\bullet}$, for $k=1,2,\ldots,N_r$. In this way, the collection of limit points for each sub-network $r$ will be
\be
\sw^{\bullet}_r=\ba{cc}w_{r,1}^{\bullet}\\
w_{r,2}^{\bullet}\\\vdots\\w_{r,N_r}^{\bullet}\ea\;\;\;(\mbox{\rm sub-network $r$})
\ee
and the collection of limit points for all $R$ sub-networks is
\be
\sw^{\bullet}\define \ba{c}\sw_1^{\bullet}\\\hline
\sw_2^{\bullet}\\\hline
\vdots\\\hline
\sw_R^{\bullet}
\ea
\ee
The arguments that follow will allow us to identify the limit vector $\sw^{\bullet}$ for the sub-networks from group $R$ in terms of the Pareto solutions of the sub-networks from group $S$. Specifically, it will hold that
\be
	\sw^{\bullet}= \mathcal{W}\tran \sw^{\star}
\label{kajdk1.3.13lk}\ee

%

\noindent where $\mathcal{W} = W \otimes I_M$ and $W$ is the matrix transformation defined by (\ref{eq:defW}). Result (\ref{kajdk1.3.13lk}) is established further ahead in (\ref{kakd13}) by showing that each agent in the network approaches the corresponding limit point in (\ref{kajdh6173.1l3k}) below within a mean-square-error that is in the order of $\mu_{\max}$. 

Now recall that each column of $W$ adds up to one, which implies that if all $S$ sub-networks happen to have the same limit point, say, $w^{\star}$, then it would follow from (\ref{kajdk1.3.13lk}) that  all agents in group $R$ will also converge to this same limit point. More generally,  if  sub-networks within group $S$ have different limit points, $w^{\star}_s$, then each agent in group $R$ will usually converge towards a different limit point, $w^{\bullet}_{r,k}$. One important conclusion that readily follows from (\ref{kajdk1.3.13lk}) is that the limit points of all agents in group $R$ are fully determined by the limit points of group $S$.

We collect the limiting points for all agents across the network into the vector
\be
	\sw_{\infty} \stackrel{\triangle}{=}
	\left[
		\begin{array} {c}
			\sw^{\star} \\\hline
			\sw^{\bullet} \\
		\end{array}
	\right]
	\begin{array} {c}
		\longrightarrow (\mbox{\rm for group $S$})\\
		\longrightarrow (\mbox{\rm for group $R$})\\
	\end{array}
\label{kajdh6173.1l3k}\end{equation}
Before justifying (\ref{kajdk1.3.13lk}), we establish a crucial property for $\sw_{\infty}$, and which will play a key role in the convergence analysis in future sections.

\begin{lemma}[{\sc Relating $\sw_{\infty}$ to ${\cal A}$}] Introduce the extended combination policy $\mathcal{A} = A \otimes I_M$. Then, it holds that $\sw_{\infty}$ is a right-eigenvector for ${\cal A}\tran$ corresponding to the eigenvalue at one, namely,
\begin{equation}
	\sw_\infty = \mathcal{A}\tran \sw_\infty \label{label.w_and_A}
\end{equation}
Eigenvectors are non-unique.
However, if we fix the top entries of the right-eigenvector of ${\cal A}\tran$ to  $\sw^{\star}$, then the solution to the linear system of equations $x={\cal A}\tran x$ is unique and equal to $\sw_{\infty}$.
\end{lemma}

\bp In view of (\ref{eq.thisx}), establishing (\ref{label.w_and_A}) is equivalent to establishing the validity of the following identity:
\begin{equation}
	\left[
		\begin{array} {c}
			\sw^{\star} \\
			\sw^{\bullet} \\
		\end{array}
	\right]
	=
	\left[
		\begin{array} {cc}
			\mathcal{T}\tran_{SS} & 0\\
			\mathcal{T}\tran_{SR} & \mathcal{T}\tran_{RR}\\
		\end{array}
	\right]
	\left[
		\begin{array} {c}
			\sw^{\star} \\
			\sw^{\bullet} \\
		\end{array}
	\right]
\label{top.row}\end{equation}
where $\mathcal{T}_{SS}  = T_{SS} \otimes I_M,$ $ \mathcal{T}_{SR} =T_{SR} \otimes I_M$, and $\mathcal{T}_{RR}  = T_{RR} \otimes I_M$.

Let ${\cal A}_s=A_s\otimes I_M$ for each sub-network $s$ from group $S$. The top row of (\ref{top.row}) is obviously true since the blocks on the diagonal of ${\cal T}_{SS}\tran$ are the matrices $\{{\cal A}_s\tran\}$ and each ${\cal A}_s$ is left-stochastic. Thus, using properties of Kronecker products, note that
\bq
\mathcal{T}\tran_{SS}\sw^{\star}&=&\mbox{\rm blockdiag}\left\{{\cal A}_s\tran(\one_{N_s}\otimes w_s^{\star})\right\}\nn\\
&=&\mbox{\rm blockdiag}\{(A_s\tran\otimes I_M)(\one_{N_s}\otimes w_s^{\star})\}\nn\\
&=&\mbox{\rm blockdiag}\{(A_s\tran\one_{N_s}\otimes w_s^{\star})\}\nn\\
&=&\mbox{\rm blockdiag}\{\one_{N_s}\otimes w_s^{\star}\}\nn\\
&=&\sw^{\star}
\eq
as claimed.

Next, observe that the second block row in (\ref{top.row}) is equivalent to requiring the following equivalent conditions:
\bq
\sw^{\bullet} = \mathcal{T}\tran_{SR} \sw^{\star} +   \mathcal{T}\tran_{RR}  \sw^{\bullet}&\Longleftrightarrow &  \mathcal{T}\tran_{SR} \sw^{\star} =  \left(I -  \mathcal{T}\tran_{RR} \right){\sw}^{\bullet} \nn\\
	&\stackrel{(\ref{kajdk1.3.13lk})}{\Longleftrightarrow}&  \mathcal{T}\tran_{SR} {\sw}^{\star}  =  \left(I -  \mathcal{T}\tran_{RR} \right)\mathcal{W}\tran {\sw}^{\star}  \nn\\
	&\stackrel{(a)} {\Longleftrightarrow}&  \mathcal{T}\tran_{SR} {\omega}^{\star}  =   \mathcal{T}\tran_{SR}  {\omega}^{\star}
\eq
where step (a) is because of equation (\ref{eq:defW}). The last condition is a trivial equality and, therefore, we conclude that relation (\ref{label.w_and_A}) holds as a result of (\ref{kajdk1.3.13lk}).

We now examine the uniqueness of the solution to (\ref{label.w_and_A}). Suppose there exists another solution, $\sw_\infty'$, satisfying (\ref{label.w_and_A}) with the same top entry $\sw^{\star}$ but with a possibly different bottom entry, denoted by $\sw^{\circ}$. Then, $\sw^{\circ}$ must satisfy:
\be
	\sw^{\circ} = \mathcal{T}\tran_{SR} \sw^{\star} +   \mathcal{T}\tran_{RR}  \sw^{\circ} \nn
\ee
Subtracting this equation from the same equation satisfied by $\sw^{\bullet}$, we find that
\be
(I-\mathcal{T}\tran_{RR}) \left( \sw^{\circ} -  \sw^{\bullet} \right)\;=\;0
\ee
However, we know from the proof of Lemma~\ref{kajd713.lemma} that  $\rho(\mathcal{T}\tran_{RR})<1$ so that $(I-\mathcal{T}\tran_{RR})$ is nonsingular. We conclude that $\sw^{\circ}=\sw^{\bullet}$, as claimed.

\ep

\section{Stability of Error Moments}
For a generic agent $k$, we let $w_{k}^{\star}$ denote its limit point. We know from the discussion in the previous section that the value of this limit point will depend on whether agent $k$ belongs to group $S$ or group $R$. Specifically, we have that
\be
w_{k}^{\star}\;=\;\left\{\begin{array}{ll}w_{s}^{\star},&(\mbox{\rm when $k\in$ sub-network $s$ in group $S$})\\
w_{r,k}^{\bullet},&(\mbox{\rm when $k\in$ sub-network $r$ in group $R$})
\end{array}\right.
\ee
We denote  the error vectors relative to this limit point by:
\bq
	 \widetilde{\bm{\psi}} _{k,i}  	& \define & w _{k}^{\star}  - \bm{\psi} _{k,i} \\
	 \widetilde{\w}_{k,i} 		& \define & w_{k}^{\star} - \w_{k,i}
\eq
If we now subtract $w_{k}^{\star}$ from both sides of the diffusion strategy (\ref{label.eq4a})--(\ref{label.eq4b}), we get
\bq
	\widetilde{\bm{\psi}} _{k,i}  &=& \widetilde{\w}_{k,i-1} + \mu_{k} \,\widehat{\nabla_{w\tran} J}_{k}(\w_{k,i-1})  \label{label.eq39a}\\
	\widetilde{\w}_{k,i}  &=& w_{k}^{\star} -  \displaystyle \sum_{\ell \in \mathcal{N}_{k}} a_{\ell k}\; \bm{\psi}_{\ell,i} \label{label.eq39b}
\eq
It follows from property  (\ref{label.w_and_A}) that the limit point $w_{k}^{\star}$ satisfies:
\be
	w_{k}^{\star} = \sum_{\ell\in {\cal N}_k} a_{\ell k} w_{\ell}^{\star}
\ee
This result allows us to absorb $w_{k}^{\star}$ into the right-most term in (\ref{label.eq39b}) so that
\be
	\widetilde{\w}_{k,i}  = \displaystyle \sum_{\ell \in \mathcal{N}_{k}} a_{\ell k}\; \widetilde{\bm{\psi}}_{\ell,i} \label{eq.eq49}
\ee
To proceed, we recall the definition of the gradient noise process from (\ref{label.eqRR}) and introduce the following block quantities, which collect variables from across all agents in the network (from within both groups $S$ and $R$ of sub-networks; the first group has a total of $N_{gS}$ agents, numbered $1$ through $N_{gS}$, and both groups have a total of $N$ agents):
\bq
\widetilde{\swb}_{S,i} &\define &
\ba{c}\widetilde{\w}_{1,i} \\ \widetilde{\w}_{2,i}\\ \vdots \\ \widetilde{\w}_{N_{gS},i}\ea \;\;\;\;\;\;(\mbox{\rm group $S$})\\
\widetilde{\swb}_{R,i} &\define &
\ba{c}\widetilde{\w}_{N_{gS}+1,i} \\ \widetilde{\w}_{N_{gS+2},i}\\ \vdots \\ \widetilde{\w}_{N,i}\ea \;\;\;(\mbox{\rm group $R$})\\
\H_{k,i} &\define & \int_0^1 \nabla_{w\tran}^2 J_{k}\left(w_{k}^{\star} - t \widetilde{\bm{\w}} _{k,i}\right)) \mathrm{d}t  \label{eq.eq50}\\
\bm{\cal H}_{S,i} &\define& \mathrm{diag}\{\H_{1,i},\H_{2,i},\cdots, \H_{N_{gS},i}\}\label{eq.eq51}\\
\bm{\cal H}_{R,i} &\define& \mathrm{diag}\{\H_{N_{gS}+1,i},\cdots, \H_{N,i}\}  \label{eq.eq52}\\
b_{S} &\define& -\ba{c}\nabla_{w\tran} J_{1}(w_{1}^{\star}) \\ \vdots \\ \nabla_{w\tran} J_{N_{gS}}(w_{N_{gS}}^{\star})\ea \\
b_{R} &\define& -\ba{c}\nabla_{w\tran} J_{N_{gS}+1}(w_{N_{gS}+1}^{\star}) \\ \vdots \\ \nabla_{w\tran} J_{N}(w_{N}^{\star})\ea \\
\mathcal{M}_{S} &\define& \mathrm{diag} \{\mu_{1}I_M,\mu_{2}I_M,\ldots,\mu_{N_{gS}}I_M \}\\
\mathcal{M}_{R} &\define& \mathrm{diag} \{\mu_{N_{gS}+1}I_M,\ldots,\mu_{N}I_M \}\\
\ssb_{S,i} &\define& \ba{c}\s_{1,i} \\ \s_{2,i}\\ \vdots \\ \s_{N_{gS},i}\ea,\;\;\ssb_{R,i} \define \ba{c}\s_{N_{gS}+1,i} \\ \s_{N_{gS}+2,i}\\ \vdots \\ \s_{N,i}\ea
\eq

Using the mean-value theorem \cite{Sayed14}, it holds that:
\bq	 \nabla_{w\tran} J_{k}(\w_{k,i}) &=& \nabla_{w\tran} J_{k}(w_{k}^{\star}) - \\
&&\left[ \int_0^1 \nabla_{w\tran}^2 J_{k}(w_{k}^{\star} - t \widetilde{\w} _{k,i}) dt \right] \widetilde{\w} _{k,i}\nn
\eq
Substituting (\ref{label.eq39a}) and (\ref{eq.eq49}), we find that the error dynamics across the network evolves according to the following recursion:
\be
\ba{c}\widetilde{\swb}_{S,i} \\
\widetilde{\swb}_{R,i} \ea
=
{\cal A}\tran\ba{c}
\mathcal{M}_{S}\cdot \ssb_{S,i} \\
\mathcal{M}_{R}\cdot \ssb_{R,i}
\ea\;	-\;
{\cal A}\tran\ba{c}
\mathcal{M}_{S}\cdot b_{S} \\
\mathcal{M}_{R}\cdot b_{R}
\ea \hspace{0.1cm}\label{eq:ErrRec}
\ee
\be
\hspace{1cm}{}+
{\cal A}\tran
\left(
I
-
\ba{c@{\hspace{-0.3cm}}c}
\mathcal{M}_{S} \bm{\cal H}_{S,i-1} & \\
& \mathcal{M}_{R} \bm{\cal H}_{R,i-1}
\ea
\right)
\ba{cc}
\widetilde{\swb}_{S,i-1} \\
\widetilde{\swb}_{R,i-1}
\ea
\nn
\ee
The next statement establishes that the entries in (\ref{kajdh6173.1l3k}) are indeed the limit points of the agents in the network in that the agents converge towards them in the mean-square error sense.

\begin{theorem}[{\sc Mean-Square-Error Stability}]
Assume the cost functions $J_k(w)$ are convex, for all agents $k=1,\ldots,N$, and that
 at least one of the cost functions in each of the $S$ sub-networks is strongly convex.
 Assume also that the first and second-order moments of the gradient noise process (\ref{label.eqRR})
 satisfy the conditions stated in Assumption 8.1 from \cite{Sayed14}. Then, the weakly-connected network is mean-square stable for sufficiently small step-sizes, namely, for every $k$ it holds that,
\begin{equation}
	\limsup_{i\to \infty} \Ex \|\widetilde{\w}_{k,i} \|^2 = O(\mu_\mathrm{max})
\label{kakd13}\end{equation}
\label{theorem.1}
\end{theorem}
\bp Using the block-triangular structure of $A$ from (\ref{eq.thisx}), the error
recursion (\ref{eq:ErrRec}) decouples into two separate recursions with the evolution of the error vectors in group $S$ being independent of the evolution of the error vectors in group $R$. In particular, it holds that
\be
	\widetilde{\swb}_{S,i}={\cal T}_{SS}\tran (I-{\cal M}_{S}\bm{\cal H}_{S,i-1})	 \widetilde{\swb}_{S,i-1}+
{\cal T}_{SS}\tran{\cal M}_s\ssb_{S,i}-{\cal T}_{SS}\tran{\cal M}_S b_S
\label{skad681312}\ee
where ${\cal T}_{SS}$ is left-stochastic. This recursion is a special case of the form studied by Theorem 9.1 in \cite{Sayed14}. Therefore, for agents in group $S$, it follows from the result of that theorem that (\ref{kakd13}) holds. Moreover, the arguments used in the proof of the theorem establish that construction (\ref{kajdj671381.,1}) indeed corresponds to the limiting points for the sub-networks in group $S$.

We still need to examine the stability of the agents from group $R$. For these agents, we conclude from (\ref{eq:ErrRec})  that the error recursion evolves instead as follows:
\bq
\widetilde{\swb}_{R,i} &=&  \mathcal{T}\tran_{RR}\left(I -\mathcal{M}_{R} \bm{\cal H}_{R,i-1}  \right)  	 \widetilde{\swb}_{R,i-1} + \nn\\
&&		\mathcal{T}\tran_{SR} \left(I -\mathcal{M}_{S} \bm{\cal H}_{S,i-1}  \right)   	 \widetilde{\swb}_{S,i-1} + \nn\\
&&\mathcal{T}\tran_{SR}\mathcal{M}_{S} \ssb_{S,i} +\mathcal{T}\tran_{RR}\mathcal{M}_{R} \ssb_{R,i} - \nn\\
&&		\mathcal{T}\tran_{SR}\mathcal{M}_{S} b_{S} - \mathcal{T}\tran_{RR}\mathcal{M}_{R}  b_{R}
\label{label.eq47}
\eq
We introduce the Jordan canonical decomposition of the matrix $T_{RR}$ as:
\be
	T_{RR} = V_{\epsilon}J_{\epsilon}V_{\epsilon}^{-1} \label{eq.63}
\ee
where $J_{\epsilon}$ consists of Jordan blocks except that the ones on the lower diagonal are replaced by a small positive scalar $\epsilon$ \cite{Horn03}. We recall from the proof of Lemma~\ref{kajd713.lemma} that $\rho(J_{\epsilon})<1$. We also introduce the extended version of (\ref{eq.63}):
\be
	\mathcal{T}_{RR} = \mathcal{V_{\epsilon}J_{\epsilon}V_{\epsilon}}^{-1}
\ee
where $ \mathcal{V_{\epsilon} }=V_{\epsilon}\otimes I_M$ and $ \mathcal{J_{\epsilon} }=J_{\epsilon}\otimes I_M$, and define the transformed quantities
\be
	\bar{\swb}_{R,i}=\mathcal{V}_{\epsilon}\tran\widetilde{\swb}_{R,i},\quad
	\bar{\mathcal{T}}\tran_{SR}= \mathcal{V}_{\epsilon}\tran\mathcal{T}\tran_{SR},\quad \bar{\mathcal{T}}\tran_{RR} \define \mathcal{V}\tran_{\epsilon}\mathcal{T}\tran_{RR} \label{eq.def67}
\ee
If we now multiply both sides of (\ref{label.eq47}) from the left by $\mathcal{V}_{\epsilon}\tran$ we obtain the equivalent recursion:
\bq
\bar{\swb}_{R,i} &=&  \mathcal{J}_{\epsilon}\tran\left(I - \mathcal{V}_{\epsilon}\tran\mathcal{M}_{R}\bm{\cal H}_{R,i-1}  \mathcal{V}_{\epsilon}^{-\mathsf{T}}  \right)  \bar{\swb}_{R,i-1} + \nn\\
&&		\bar{\mathcal{T}}\tran_{SR} \left(I -\mathcal{M}_{S} \bm{\cal H}_{S,i-1}  \right)   \widetilde{\swb}_{S,i-1} + \nn\\
&&	\bar{\mathcal{T}}\tran_{SR}\mathcal{M}_{S} \ssb_{S,i} +\bar{\mathcal{T}}\tran_{RR}\mathcal{M}_{R} \ssb_{R,i} - \nn\\
&&		\bar{\mathcal{T}}\tran_{SR}\mathcal{M}_{S} b_{S} - \bar{\mathcal{T}}\tran_{RR}\mathcal{M}_{R}  b_{R}
\label{label.eq49}
\eq
In Appendix~\ref{app.a} we start from this recursion and complete the argument to conclude that  (\ref{kakd13}) also holds for all agents in group $R$.

\ep

Besides the stability of the second-order moment of the network error dynamics, we can similarly establish the stability of the first and fourth-order moments of erros. These results are useful in evaluating the performance of the network in the next section.

\begin{theorem}[{\sc Error Moment Stability}]
Assume the cost functions $J_k(w)$ are convex, for all agents $k=1,\ldots,N$, and that
 at least one of the cost functions in each of the $S$ sub-networks is strongly convex.
 Assume also that the first and fourth-order moments of the gradient noise process  (\ref{label.eqRR})
 satisfy the conditions stated in Theorem 9.2 from \cite{Sayed14}. Assume further that the individual costs $J_k(w)$ satisfy the following smoothness condition relative to their limits points:
 \be
 \|\nabla_w^2\,J_k(w_k^{\star}+\Delta w)-\nabla_w^2\,J_k(w^{\star})\|\;\leq\;\kappa_d\|\Delta w\|
\label{jakd.187312} \ee
 for small perturbations $\|\Delta w\|\leq\epsilon$ and for some $\kappa_d\geq 0$.  Then, for sufficient small step-sizes, it holds that:
\bq
	\limsup_{i\to \infty} \|\Ex \widetilde{\w}_{k,i} \| &= &O(\mu_\mathrm{max})\label{result.11}\\
	\limsup_{i\to \infty} \Ex \|\widetilde{\w}_{k,i} \|^4 &=& O(\mu_\mathrm{max}^2)\label{result.22}\eq
\label{theorem.2a}
\end{theorem}

\bp We again consider the error recursion (\ref{skad681312}), which is a special case of the forms studied in Theorems 9.2 and 9.6 in \cite{Sayed14}. Therefore, for agents in group $S$, it follows from the results of these theorems that results (\ref{result.11})--(\ref{result.22}) hold.

We still need to establish that  (\ref{result.11})--(\ref{result.22}) hold for all agents from group $R$. For these agents, we start from their error recursion (\ref{label.eq49}). The derivation in Appendices~\ref{app.b} and~\ref{app.c} complete the derivation and show that (\ref{result.11})--(\ref{result.22}) hold for these agents as well.
\ep

\section{Long-Term Network Dynamics}
We can be more specific and assess the size of the mean-square deviation (MSD) defined by (\ref{kakd13})  to first-order in $\mu_{\max}$. If we examine recursion (\ref{eq:ErrRec}), we find that it is a nonlinear and stochastic difference recursion due to the dependence of the Hessian matrices $\bm{\cal H}_{S,i}$ and $\bm{\cal H}_{R,i}$ on the error vectors $\widetilde{\swb}_{S,i}$ and  $\widetilde{\swb}_{R,i}$. To address this difficulty and arrive at closed-form expressions for the MSD levels at the various agents, we replace (\ref{eq:ErrRec}) by  a long-term model that will be shown to provide an accurate approximation for the behavior of the network as $i\to \infty$ and for small step-sizes.  

For this purpose, we extend a construction from \cite[Ch. 10]{Sayed14} to the current setting and introduce the constant Hessian matrices:
\bq
	 {\cal {H}}_{S} &\define& {\rm diag} \{ H_1,H_2,\ldots, H_{N_{gS}}\}  \label{eq.consHS}\\
	 {\cal {H}}_{R} &\define& {\rm diag} \{ H_{N_{gS}+1},H_{N_{gS}+2},\ldots, H_{N}\} \label{eq.consHR}\\
	 H_k                &\define& \nabla_w^2 J_k(w^{\star}_k)
\eq
We also define the error matrices:
\bq
	\bm{\cal \widetilde{H}}_{S,i} &\define& {\cal {H}}_{S}  - \bm {\cal {H}}_{S,i} \label{eq.longHerrS}\\
	\bm{\cal \widetilde{H}}_{R,i} &\define& {\cal {H}}_{R}  - \bm {\cal {H}}_{R,i} \label{eq.longHerrR}
\eq
We then replace the original error-recursion (\ref{eq:ErrRec}) by the following model where the Hessian matrices $\bm{\cal H}_{S,i-1}$ and $\bm{\cal H}_{R,i-1}$ are replaced by the constant values (\ref{eq.consHS})--(\ref{eq.consHR}): \vspace{0.2cm}
\be
\ba{c}\widetilde{\swb}'_{S,i} \\
\widetilde{\swb}'_{R,i} \ea
=
{\cal A}\tran\ba{c}
\mathcal{M}_{S}\cdot \ssb_{S,i} \\
\mathcal{M}_{R}\cdot \ssb_{R,i}
\ea\;	-\;
{\cal A}\tran\ba{c}
\mathcal{M}_{S}\cdot b_{S} \\
\mathcal{M}_{R}\cdot b_{R}
\ea \hspace{0.1cm}\label{eq:ErrRec2}
\ee
\be
\hspace{1cm}{}+
{\cal A}\tran
\left(
I
-
\ba{c@{\hspace{-0.3cm}}c}
\mathcal{M}_{S}  {\cal H}_{S} & \\
& \mathcal{M}_{R} {\cal H}_{R}
\ea
\right)
\ba{cc}
\widetilde{\swb}'_{S,i-1} \\
\widetilde{\swb}'_{R,i-1}
\ea
\nn
\ee
Observe that in this model we are using the prime notation to refer to its state variables. The next result explains why the above model provides a good approximation for the actual network performance in steady-state. 

\begin{theorem}[{\sc Accuracy of Approximation}] 
Under the same conditions of Theorem \ref{theorem.1}, it holds that, for sufficiently small step-sizes the state variables of models (\ref{eq:ErrRec}) and (\ref{eq:ErrRec2}) are close to each other, namely, :
 \be
	 \limsup_{i\to\infty} \Ex\left \| \ba{c}\widetilde{\swb}_{S,i} \\ \widetilde{\swb}_{R,i}\ea -\ba{c}\widetilde{\swb}'_{S,i} \\ \widetilde{\swb}'_{R,i}\ea \right \|^2 = O(\mu_{\max}^2) \label{gongshi.79}\\
	 \ee
	 \be
	 \limsup_{i\to\infty} \Ex \left \|\ba{c}\widetilde{\swb}_{S,i} \\ \widetilde{\swb}_{R,i}\ea \right\|^2 =  \limsup_{i\to\infty} \Ex \left\|\ba{c}\widetilde{\swb}'_{S,i} \\ \widetilde{\swb}'_{R,i}\ea \right\|^2 + O(\mu_{\max}^{3/2}) \label{gongshi.80}
\ee
\label{theorem.longterm}
\end{theorem}
\bp 
In Theorem 10.2 of \cite{Sayed14}, it was established that the error bounds in (\ref{gongshi.79}) and (\ref{gongshi.80}) hold for strongly-connected networks and, hence, for all agents of type $S$. Appendix \ref{app.longterm} completes the argument and shows that the same conclusion holds for agents of type $R$. 
\ep

We remark that in a manner similar to the proof of Theorems \ref{theorem.1} and \ref{theorem.2a}, we can also establish that 
\bq
	\limsup_{i\to \infty} \|\Ex \widetilde{\w}'_{k,i} \| &= &O(\mu_\mathrm{max})\label{result.11long}\\
	\limsup_{i\to \infty} \Ex \|\widetilde{\w}'_{k,i} \|^2 &= &O(\mu_\mathrm{max})\label{result.00long}\\
	\limsup_{i\to \infty} \Ex \|\widetilde{\w}'_{k,i} \|^4 &=& O(\mu_\mathrm{max}^2)\label{result.22long}
\eq

\section{Mean-Square-Error Performance}
In this section, we evaluate the steady-state mean-square deviation (MSD), defined as
the value of the error variance, $\Ex\|\widetilde{\w}_{k,i}\|^2$, as $i\rightarrow\infty$
and for sufficiently small step-sizes. The following result holds depending on whether agent
$k$ belongs to group $S$ or group $R$. For agents in group $R$, we introduce the
$S\times 1$ column vector:
\begin{equation}
	c_k =
	\left[
	 \begin{array}{cccc}
		    \one\tran_{N_1} & \ & \  &  \ \\ [-2pt]
		    \ & \one\tran_{N_2} & \  & \  \\[-2pt]
		    \ & \ & \ddots & \ \\ [-2pt]
		    \ & \ & \ &  \one\tran_{N_S}\\
	 \end{array}
	\right]
	\left[W\right]_{:,k} \label{eq.definef}
\end{equation}
where the notation $[W]_{:,k}$ refers to the column of $W$ from (\ref{eq:defW}) corresponding to agent $k$. One useful interpretation for the entries of the vector $c_k$ is as follows. Since each column of $W$ adds up to one, therefore, the entries of $c_k$ will add up to one. In addition, as expression (\ref{eq.MSD2}) below reveals, the square of the $s-$th entry of
$c_k$ will measure the influence of sub-network $s$ from group $S$ on the performance of agent $k$ from group $R$.

\begin{theorem}[{\sc MSD Performance}] Assume agent $k$ belongs to a sub-network $s$ from
group $S$. Then, all agents within sub-network $s$ achieve the same MSD level given by
\begin{equation}
	\mbox{\rm MSD}_s = \frac{1}{2} \Tr   \left[\left(\sum_{k=1}^{N_s}q_{s,k}H_{s,k} \right)^{-1}\left( \sum_{k=1}^{N_s}q_{s,k}^2{G}_{s,k}\right)\right]
\label{eq.MSD1}
\end{equation}
where we are using the notation $\{H_{s,k},G_{s,k}\}$, with a subscript $s$, to denote
the Hessian and covariance matrices  (\ref{label.eq11a})--(\ref{label.eq11b}) for agent $k$
within sub-network $s$, in the same manner that we used $s$ 
in the definition of the weighting scalars
$\{q_{s,k}\}$ in (\ref{definqa.s}).

Assume, on the other hand, that agent $k$ belongs to a sub-network $r$ from group $R$. Then, in this case, its MSD level will be given by
\begin{equation}
\mbox{\rm MSD}_R(k) = \sum_{s=1}^S c_k^2(s) \cdot \mbox{\rm MSD}_{s}
\label{eq.MSD2}
\end{equation}
where $c_k(s)$ denotes the $s-$th entry of vector $c_k$. That is, the performance of
any agent from group $R$ is given by a weighted combination of the MSD performance levels
of the sub-networks from group $S$.

\label{theorem.3}
\end{theorem}

\bp The argument is given in Appendix~\ref{app.d}.

\ep

\section{Simulation Results}
We illustrate the theoretical results  by considering several examples. 

\begin{figure}[h]
\epsfxsize 8cm \epsfclipon
\begin{center}
\leavevmode \epsffile{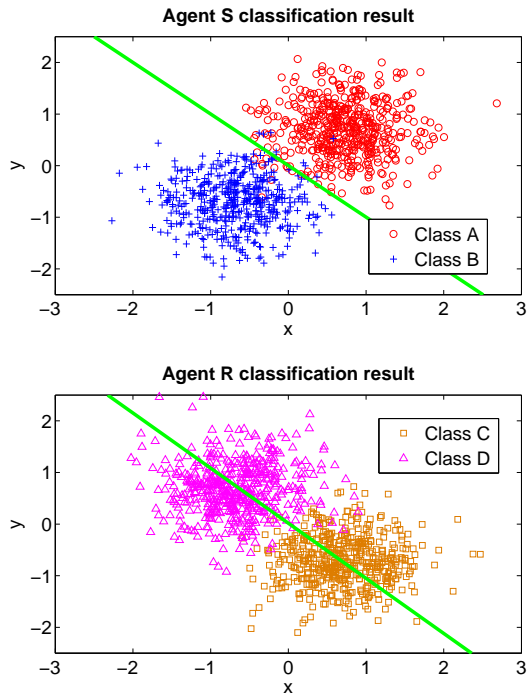} \caption{{\small  Agents $S$ and $R$ use logistic regression to learn the separating hyperplanes for their data. The agents are weakly connected. The simulation result shows that  the behavior of agent $R$ is fully  misguided by agent $S$ even though the data received by agent $R$ is sufficient for it to learn its own hyperplane.}}\label{fig-D.label}
\end{center}
\end{figure}

Our first example relates to a pattern classification problem and illustrates how a group $S$ of agents can dramatically influence the learning behavior of agents in another group  $R$. Thus, consider initially a simple network consisting of $N=2$ agents, where one node plays the role of the $S$ agent and the second node plays the role of the $R$ agent. We assume each agent is observing data that belong to different classes: data at node $S$ belong to one of two classes $\{A,B\}$, while data at node $R$ belong to one of two other classes $\{C,D\}$. If each agent were to learn independently of the other agent, then they would be able to determine their respective separating hyper-planes and classify their data with reasonable accuracy. Here, however, we would like to examine the influence of node $S$ on node $R$ when both agents are connected by means of a weak topology say, one with combination matrix --- see Fig. \ref{fig-2agents.label}: 
\begin{figure}[h]
\epsfxsize 5.0cm \epsfclipon
\begin{center}
\leavevmode \epsffile{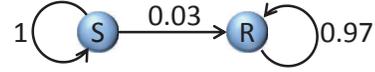}\caption{{\small An example of a weakly-connected two-agent network.}} \label{fig-2agents.label}
\end{center}
\end{figure}
\be
	A=
	\ba{cc}
		1& 0.03 \\
		0& 0.97
	\ea
\ee
This choice for $A$ means that agent $R$ is careful  in  dealing with agent $S$ and is only assigning weight $0.03$ (or $3\%$) to its interaction with $S$. Still, despite this level of caution, the corresponding matrix $W$ defined by  (\ref{eq:defW}) reduces to a scalar in this case and evaluates to \be W=0.03(1-0.97)^{-1}=1\ee In other words, agent $R$ will be completely misguided by agent $S$.  This effect is evident in Fig.~\ref{fig-D.label}, where it is seen that agent $S$ (top plot) is able to determine its separating hyperplane, while agent $R$ (bottom plot) is driven to select the same hyperplane (which is obviously wrong for its data). In this simulation, both agents are running a logistic regression classifier \cite{Sayed14,bishop2006pattern,theo2008}

We consider next a more involved example based on the topology shown earlier in Fig.~\ref{fig-C.label} to show that a weakly-connected network can sometimes achieve better performance than strongly-connected networks. The objective now is to determine an elliptical curve that separates the data into two classes: class +1 consists of data that are  concentrated inside the curve and class -1 consists of data that are concentrated outside the curve. We assume about $10\%$ of the data available to agents $R$ are outliers. The outlier data belong to class +1 but are located away from the origin.  Obviously, since we are dealing with a weakly-connected network, agents in group $S$ will not be affected by these outliers since the outliers are only sensed by the $R$ agents, which in turn do not feed information into the $S$ agents. This situation corresponds to a scenario where weak connectivity is advantageous.

\begin{figure}[h]
\epsfxsize 7.2cm \epsfclipon
\begin{center}
\leavevmode \epsffile{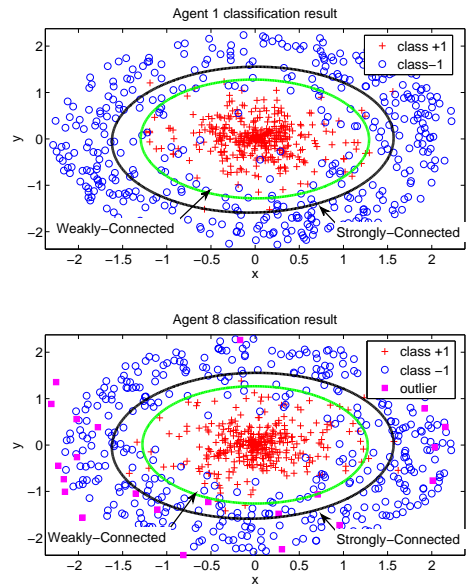} \caption{{\small Logistic classification result using an elliptic separation curve. In the simulation, agents $\{k=6,7,8\}$ in Fig. \ref{fig-C.label} suffer from outlier data. The black curve represents the result obtained by a strongly-connected network, while the green curve represents the result obtained by a weakly-connected network; observe that the curve is larger in the former case due to the influence of the outliers. }}\label{fig-D2.label}
\end{center}
\end{figure}

We  assume each agent employs the same logistic cost function:
\be
	J_k(w) =  \rho \| w \|^2\;+\;\Ex \left\{ \ln [1 + e^{-\boldsymbol{\gamma}_k \h_k\tran w}] \right\} 
\ee
where $\boldsymbol{\gamma}_k$ represents the class label $\{+1,-1\}$, $\rho$ is a regularization 
parameter, and the feature vector $\h_k$ is chosen as follows in terms of the coordinates of the measurements  \cite{bishop2006pattern,theo2008}:
\begin{align}
	\h_k(1) &= 5, & \h_k(2) &= \x, & \h_k(3) &= \y \nn \\
	\h_k(4) &= \x^2, & \h_k(5) &= \y^2, & \h_k(6) &= \x\y\nn
\end{align}
The gradient vector is estimated by using the instantaneous approximation:
\be
	\widehat{\nabla_{w\tran} J}_{k}(\w_{k,i-1}) = \displaystyle \frac{-\; \bm{\gamma}_k(i) \h_{k,i}}{1 + e^{\bm{\gamma}_k(i) \h_{k,i} \tran \w_{k,i-1}}} + \rho \w_{k,i-1}
\ee
We compare the performance of the weakly-connected topology against a strongly-connected (actually, fully-connected) network with combination matrix:
\be
	A =\frac{1}{8} \one_8 \one\tran_8 \label{network.90}
\ee
In Fig.~\ref{fig-D2.label}, we observe that the black elliptic curve, which is the result obtained by the strongly-connected network (\ref{network.90}), is larger than the green elliptic curve obtained by the weakly-connected network. Comparing both boundary curves, we  find that the black curve includes a larger proportion  of -1 data, which will be mistakenly inferred as belonging to class +1. Obviously, weakly-connected networks do not always provide
performance that is superior to fully-connected networks. There are however
situations when this can occur and this is what the simulation results are meant
to show in this example. We considered a case in which some $R-$agents are subject to outliers. In
a fully-connected network, the outliers will influence the performance because
their presence will be sensed by all agents. In a weakly-connected network, if
these outliers are affecting only the $R-$agents, then the $S-$agents will be able to
suppress their influence according to Theorem \ref{theorem.3} and performance will therefore
be improved. 

\begin{figure}[h]
\epsfxsize 8cm \epsfclipon
\begin{center}
\leavevmode \epsffile{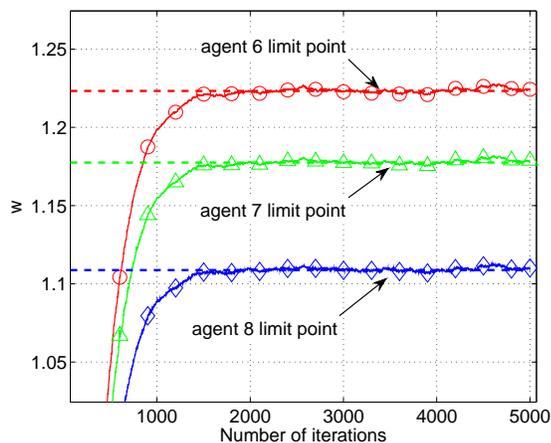} \caption{{\small Limit points for agents in group $R$.}}\label{fig-E.label}
\end{center}
\end{figure}

\begin{figure}[h]
\epsfxsize 9cm \epsfclipon
\begin{center}
\leavevmode \epsffile{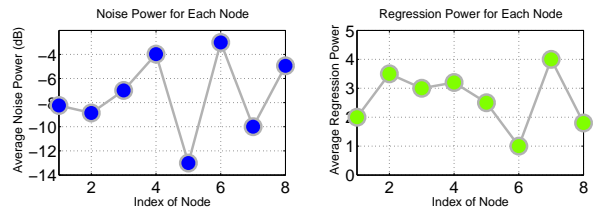} \caption{{\small Observation noise and regression power profile for all agents.}}\label{fig-G.label}
\end{center}
\end{figure}

\begin{figure}[h]
\epsfxsize 8cm \epsfclipon
\begin{center}
\leavevmode \epsffile{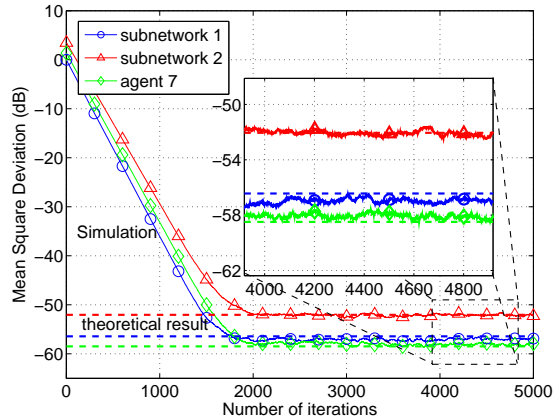} \caption{{\small  Mean-square-deviation (MSD) performance levels.}}\label{fig-F.label}
\end{center}
\end{figure}

Our third example illustrates result (\ref{kajdk1.3.13lk}) concerning limit points. We associate with each agent in Fig. \ref{fig-C.label} the quadratic cost:
\be
	J_k(w) = \Ex (\d_k(i) - \u_{k,i} w)^2
\ee
where agent $k$ senses streaming data $\{\d_k(i),\u_{k,i}\}$; the data are related to some unknown model $w^{\circ}_{k}$  via the linear  regression model:
\be
	\d_k(i) = \u_{k,i} w^{\circ}_{k} + \v_k(i)
\ee
where $\v_k(i)$ is observation noise and assumed to be spatially and temporally white.  It is sufficient for this example to assume one-dimensional limit points and to set $w^{\circ}_{k} =1$ for the agents in sub-network $1$,  
$w^{\circ}_{k} =1.5$ for the agents in sub-network $2$, and $w_k^o=1.25$ for the agents in sub-network $3$. It is obvious that the Pareto optimal solutions of the two $S$ sub-networks will be $w^{\star}_1 = 1$ and  $w^{\star}_2 = 1.5$, respectively. Computing the corresponding matrix $W$ from (\ref{label.eq11}) we get:
\be
W=
	\ba{ccc}
		0       &  0  &      0\\
   		0.4046   & 0.5267   & 0.7099 \\
   		0.1489   &  0.1183  &  0.0725 \\ \hline	
    		0.4466   & 0.3550   & 0.2176 \\
         	0  &       0  &       0
	\ea
	\hspace{-0.3cm} \label{eq.exampleW}
\begin{array}{l}
\left. \rule{0mm}{6.7mm}\right\} {\mathrm{subnetwork\ 1} }\\ \\
\left. \rule{0mm}{4.3mm}\right\} {\mathrm{subnetwork\ 2} }\\	
\end{array}
\ee
\noindent so that the limit points for the $R$ agents are 
\be w^{\bullet}_{r,6} = 1.2233,\;\; w^{\bullet}_{r,7} = 1.1775,\;\;
w^{\bullet}_{r,8} = 1.1088\ee
These values are consistent with the simulation results in Fig.~\ref{fig-E.label}. We further illustrate in Fig.~\ref{fig-F.label} the MSD performance for the agents of type $S$ and $R$. Figure~\ref{fig-G.label} shows the observation noise and regression power profile across the agents. In these simulations, we employed Gaussian distributed data to generate the observation noise and regression data. The step-size
parameters were set to $\mu_k=0.0005$ for all agents.  The theoretical MSD value for agents in sub-network $1$ follows from (\ref{eq.MSD1}) and is found to be $-56.43$dB. Likewise, the theoretical MSD value for agents in sub-network $2$ is found to be $-52.05$ dB. On the other hand, using (\ref{eq.exampleW}), we get
 \bq
 	c_7(1) &=& 0+ 0.5267 + 0.1183= 0.6450 \\
	c_7(2) &=& 0.3550 + 0 =0.3550
\eq
and the theoretical MSD value for agent $7$ in sub-network 3 is then found from (\ref{eq.MSD2}) to be
\begin{align}
	\mbox{\rm MSD}_R(7) &= 0.6450^2\times \mathrm{MSD}_{1} + 0.3550^2 \times \mathrm{MSD}_{2} \nn \\ 
	& = -58.49 \mathrm{dB}
\end{align}
which is consistent with the simulated result.

\section{Concluding Remarks}
In this article we examined the learning behavior of adaptive agents over weakly-connected networks. We showed that  a leader-follower relationship develops among the agents, and identified the limit points towards which the individual agents converge. We also derived closed-form expressions to quantify how close the agents get to their limit points in the mean-square-error sense, and revealed how the MSD performance of follower agents is determined by the MSD performance of leader agents. We illustrated the results by considering examples dealing with pattern classification and distributed estimation. We also illustrated how weak topologies can be beneficial in reducing the impact of outlier data.

\appendices
\section{Mean-Square-Error Stability of (\ref{label.eq49})}\label{app.a}

\noindent If we equate the squared-norm of both sides of
(\ref{label.eq49}), and compute expectations, we get
\be
\nn \Ex \|\bar{\swb}_{R,i} \|^2 \quad \quad \quad \quad \quad \quad \quad \quad \quad \quad \quad \quad \quad \quad \quad \quad \quad \quad \quad\quad
\ee
\begin{eqnarray}	
	&\stackrel{\rm (a)}{=}&
	\Ex \left\| \mathcal{J}_{\epsilon}\tran \left(I -\mathcal{V}_{\epsilon}\tran\mathcal{M}_{R} \bm{\cal H}_{R,i-1}
	\mathcal{V}_{\epsilon}^{-\mathsf{T}} \right)  \bar{\swb}_{R,i-1}\right. + \nn \\
	& & \;\;\;\;\; \bar{\mathcal{T}}\tran_{SR}\left( I -\mathcal{M}_{S} \bm{\cal H}_{S,i-1}  \right)   \widetilde{\swb}_{S,i-1} - \nn\\
	& &\;\;\;\;\; \left. \bar{\mathcal{T}}\tran_{SR}\mathcal{M}_{S} b_{S} - \bar{\mathcal{T}}\tran_{RR}\mathcal{M}_{R}  b_{R}  \right\|^2 + \nn\\
	& & \;  \Ex \|\bar{\mathcal{T}}\tran_{SR} \mathcal{M}_{S} \ssb_{S,i} +\bar{\mathcal{T}}\tran_{RR}\mathcal{M}_{R} \ssb_{R,i} \|^2  \nn \\\nn\\
	&=&
	\Ex \left\| \frac{1-t}{1-t}\mathcal{J}_{\epsilon}\tran \left(I -\mathcal{V}_{\epsilon}\tran\mathcal{M}_{R} \bm{\cal H}_{R,i-1}
	\mathcal{V}_{\epsilon}^{-\mathsf{T}} \right)  \bar{\swb}_{R,i-1}\right. + \nn \\
	& & \;\;\;\;\; t\frac{1}{t} \Big\{\bar{\mathcal{T}}\tran_{SR}\left( I -\mathcal{M}_{S} \bm{\cal H}_{S,i-1}  \right)   \widetilde{\swb}_{S,i-1} - \nn\\
	& &\;\;\;\;\; \left. \bar{\mathcal{T}}\tran_{SR}\mathcal{M}_{S} b_{S} - \bar{\mathcal{T}}\tran_{RR}\mathcal{M}_{R}  b_{R}  \Big\} \right\|^2 + \nn\\
	& & \;  \Ex \|\bar{\mathcal{T}}\tran_{SR} \mathcal{M}_{S} \ssb_{S,i} +\bar{\mathcal{T}}\tran_{RR}\mathcal{M}_{R} \ssb_{R,i} \|^2  \nn \\\nn\\
	& \stackrel{\rm (b)}{\leq} &
	\frac{1}{1-t} \Ex \left\| \mathcal{J}_{\epsilon}\tran \left(I -\mathcal{V}_{\epsilon}\tran\mathcal{M}_{R} \bm{\cal H}_{R,i-1} \mathcal{V}_{\epsilon}^{-\mathsf{T}} \right) \bar{\swb}_{R,i-1}\right\|^2 + \nn \\
	& & \frac{2}{t} \Ex \left\|  \bar{\mathcal{T}}\tran_{SR}  \left( I-\mathcal{M}_{S} \bm{\cal H}_{S,i-1}  \right) \widetilde{\swb}_{S,i-1} \right\|^2 + \nn \\
	& & \frac{2}{t} \Ex \|\bar{\mathcal{T}}\tran_{SR}\mathcal{M}_{S} b_{S} + \bar{\mathcal{T}}\tran_{RR}\mathcal{M}_{R}  b_{R} \|^2  \Big) + \nn \\
	& & 2\mu_\mathrm{max}^2 \left( \| \bar{\mathcal{T}}\tran_{SR} \|^2 \Ex  \|\ssb_{S,i} \|^2  + \|\bar{\mathcal{T}}\tran_{RR} \|^2 \Ex \|\ssb_{R,i} \|^2  \right) \nn\\\nn\\
	& \stackrel{\rm (c)}{\leq}&
	\frac{\rho(\mathcal{J_{\epsilon}J_{\epsilon}}\tran)}{1-t} \Ex\| \bar{\swb}_{R,i-1}\|^2
	+ \frac{2\| \bar{T}_{SR} \|^2}{t} \Ex \| \widetilde{\swb}_{S,i-1} \|^2 +\nn \\
	& & \frac{4}{t}\mu_\mathrm{max}^2\left( \|\bar{T}_{SR}\|^2 \|b_{S}\|^2 + \|\bar{T}_{RR}\|^2\|b_{R}\|^2  \right) + \nn \\
	& & 2\mu_\mathrm{max}^2 \left( \|\bar{T}_{SR} \|^2 \Ex  \|\ssb_{S,i} \|^2  +
	\|\bar{T}_{RR}\|^2 \Ex \|\ssb_{R,i} \|^2  \right)\label{sunc.lakdq}
\end{eqnarray}
for any $0<t<1$, where step (a) is because the gradient noise process is zero-mean and independent of all other
random variables conditioned on $\bm{\cal F}_{i-1}$, step (b) is because of Jensen's inequality
and the sub-multiplicative property of norms,
and in step (c) we used the equalities \be \|\bar{\mathcal{T}}_{RR}\| = \|\bar{T}_{RR}\|,
\;\;\;\;\|\bar{\mathcal{T}}_{12}\| = \|\bar{T}_{12}\|\ee
as well as
\bq \| I -\mathcal{V}_{\epsilon}\tran\mathcal{M}_{R}
\bm{\cal H}_{R,i-1} \mathcal{V}_{\epsilon}^{-\mathsf{T}}\| &\leq& 1 \label{eq.EQ74}\\
\| I -\mathcal{M}_{S} \bm{\cal H}_{S,i-1}\|&\leq &1 \label{eq.EQ75}
\eq for sufficiently small  $\mu_{\max}$. In the statement of the theorem, we are assuming that
the gradient noise process satisfies the conditions stated in Assumption 8.1 in \cite{Sayed14},
from which we conclude that
\bq
\Ex  \| \ssb_{S,i} \|^2 &\leq& \beta_{S}^2\, \Ex \| \widetilde{\swb}_{S,i-1}\|^2  + \sigma_{S}^2 \label{eq.StoGradS}\\
\Ex  \| \ssb_{R,i} \|^2 &\leq& \beta_{R}^2\, \Ex \| \widetilde{\swb}_{R,i-1}\|^2  + \sigma_{R}^2 \label{eq.StoGradR}
\eq
where
\bq
\beta_{S}^2 &=&\max_{1\leq k\leq N_{gS}}\,\beta_k^2\\
\beta_{R}^2 &=&\max_{N_{gS}+1\leq k\leq N}\,\beta_k^2\\
\sigma_{S}^2 &=& \sum_{k=1}^{N_{gS}} \sigma^2_{k} \\
\sigma_{R}^2 &= &\sum_{k=N_{gS}+1}^{N} \sigma^2_{k}
\eq
Now let $v^2 \define  \| \mathcal{V}^{-\mathsf{T}}_{\epsilon} \|^2$ and note that
\be
\Ex \| \widetilde{\swb}_{R,i-1}\|^2 = \Ex \|\mathcal{V}^{-\mathsf{T}}_{\epsilon} \bar{\swb}_{R,i-1}\|^2 \;\leq\;
v^2\,  \Ex  \|\bar{\swb}_{R,i-1}\|^2 \label{eq.wBarTilde}
\ee
Then, we can write
\bq
\Ex  \| \ssb_{R,i} \|^2 &\leq& v^2\beta_{R}^2\, \Ex \| \widetilde{\swb}_{R,i-1}\|^2  + \sigma_{R}^2
\eq
Substituting these expressions into (\ref{sunc.lakdq}) gives:

\be
\hspace{-6cm} \displaystyle \Ex \|\bar{\swb}_{R,i} \|^2 \nn \vspace{-0.3cm}
\ee
\bq
&\leq& \nn
\left( \frac{\rho(\mathcal{J_{\epsilon}J_{\epsilon}}\tran)}{1-t} + 2\mu_{\max}^2\|\bar{T}_{RR}\|^2
v^2\beta_{R}^2 \right) \Ex \| \bar{\swb}_{R,i-1}\|^2  + \\ \nn
& &  \left( \frac{2\|\bar{T}_{SR}\|^2}{t} + 2\mu_{\max}^2\|\bar{T}_{SR}\|^2
\beta_{S}^2 \right) \Ex \| \widetilde{\swb}_{S,i-1}\|^2  + \\ \nn
& & \frac{4}{t}\mu_\mathrm{max}^2\left( \|\bar{T}_{RR}\|^2\|b_{R}\|^2 + \|\bar{T}_{SR}\|^2 \|b_{S}\|^2 \right) + \\
& & 2\mu_\mathrm{max}^2 \left( \|\bar{T}_{RR}\|^2\sigma_{R}^2 + \|\bar{T}_{SR}\|^2 \sigma_{S}^2 \right) \label{eq.1}
\eq
For compactness of notation, we introduce the scalar coefficients:
\bq
\sigma_{RR}^2 &=&\|\bar{T}_{RR}\|^2\\
\sigma_{SR}^2 &=& \|\bar{T}_{SR}\|^2  \label{eq.scalar1}\\
b^2 &=& \|\bar{T}_{RR}\|^2 \|b_{R}\|^2 + \|T_{SR}\|^2 \|b_{S}\|^2  = O(1)    			 \label{eq.ScalarB}\\
c^2 &=& \|\bar{T}_{RR}\|^2 \sigma_{R}^2 + \|T_{SR}\|^2 \sigma_{S}^2 = O(1)		 \label{eq.ScalarC}
\eq

Before proceeding, let us examine the spectral radius of the matrix ${\cal J}_{\epsilon}{\cal J}_{\epsilon}\tran$. For that purpose, we note that
\be
\rho(\mathcal{J_{\epsilon}J_{\epsilon}}\tran) = \rho(J_{\epsilon}J_{\epsilon}\tran)\leq\|J_{\epsilon}J_{\epsilon}\tran\|_1
\ee
where the last inequality is because the spectral radius of a matrix is bounded by any of its norms; the last norm is the one-norm (maximum absolute column sum). We next verify that
\be
\|J_{\epsilon}J_{\epsilon}\tran\|_1\;=\;\rho^2(T_{RR})
\label{kajd76813.lk12lk3}\ee
so that
\be
\rho(\mathcal{J_{\epsilon}J_{\epsilon}}\tran)\;\leq\;\rho^2(T_{RR})\label{usiadk..1lk3l1k2}
\ee

To establish (\ref{kajd76813.lk12lk3}), we proceed as follows. The eigenvalues of the matrix $T_{RR}$ are composed of the eigenvalues of the matrices $\{A_k, k= S+1,\cdots, S+R\}$ that appear on its block diagonal. Each of these matrices is associated with a connected network and has nonnegative entries. It follows from the Perron-Frobenius Theorem \cite{Horn03,Pillai05,Meyer00} that the largest eigenvalue of each $\{A_k, k= S+1,\cdots, S+R\}$ is simple (i.e., its multiplicity is equal to one). Some of the largest eigenvalues among the $A_k$ may coincide with each other, in which case the largest eigenvalue of $T_{RR}$ may have multiplicity larger than one. However, this largest eigenvalue of $T_{RR}$ cannot be defective, meaning that its algebraic multiplicity must coincide with its geometric multiplicity. This is because the matrices $\{A_k\}$ occur at different block locations in $T_{RR}$ and the eigenvectors corresponding to their largest eigenvalues will not coincide with each other. We therefore conclude that the largest eigenvalue of $T_{RR}$ is semi-simple (its algebraic and geometric multiplicities are equal to each other). Thus, assume we order the eigenvalues of $T_{RR}$ from largest to smallest, say, as:
\be
|\lambda_1(T_{RR})|> 
|\lambda_2(T_{RR})|\geq
|\lambda_3(T_{RR})|\geq\ldots
\ee
Then, the Jordan matrix $J_{\epsilon}$ will, for example, be of the following representative form (say, for $3$ Jordan blocks of size $3\times 3$ each):
\be
J_{\epsilon}=\mbox{\rm blockdiag}\{J_1,J_2,J_3\}
\ee
where
\bq
J_1&=&\,\mbox{\rm diag}\{\lambda_1,\lambda_1,\lambda_1\}\\
J_2&=&\ba{cccc}\lambda_2\\\epsilon&\lambda_2\\&\epsilon&\lambda_2\ea\\
J_3&=&\ba{cccc}\lambda_3\\\epsilon&\lambda_3\\&\epsilon&\lambda_3\ea
\eq
It follows that
\be
{J}_{\epsilon}{J}_{\epsilon}\tran = \mathrm{blockdiag}\{K_1,K_2,K_3\}
\ee
where, for example,
\bq
K_1&=&\,\mbox{\rm diag}\{\rho^2(T_{RR}),\rho^2(T_{RR}),\rho^2(T_{RR})\}\\
K_2&=&\ba{cccc}\lambda_2^2&\epsilon\lambda_2\\\epsilon\lambda_2&\lambda_2^2+\epsilon^2\\
&
\epsilon\lambda_2&\lambda_2^2+\epsilon^2\ea
\eq
from which we conclude that, we can choose $\epsilon$ small enough such that
\bq
\|{J}_{\epsilon}{J}_{\epsilon}\tran\|_1&=&\max\{\rho^2(T_{RR}),\;\lambda_2^2(T_{RR})+2\epsilon\lambda_2(T_{RR})+\epsilon^2\}\nn\\
&=&\max\{\rho^2(T_{RR}),\;(\lambda_2(T_{RR})+\epsilon)^2\}\nn\\
&=&\rho^2(T_{RR})
\eq
as claimed.

Now we can use inequality (\ref{usiadk..1lk3l1k2}) to replace $\rho(\mathcal{J_{\epsilon}J_{\epsilon}}\tran)$ in (\ref{eq.1}) and choose $t=1-\rho(T_{RR})$, to obtain

\be
\hspace{-6cm} \displaystyle \Ex \|\bar{\swb}_{R,i} \|^2 \;\leq \nn \vspace{-0.2cm}
\ee
\bq
&& \left(\rho(T_{RR})+2\mu_\mathrm{max}^2\sigma_{RR}^2v^2
\beta_{R}^2 \right)\Ex \|\bar{\swb}_{R,i-1}\|^2 + \nn \\
& & \left( \frac{2\sigma_{SR}^2}{1 - \rho(T_{RR})} + 2\mu_\mathrm{max}^2\sigma_{SR}^2\beta_{S}^2\right)\Ex \|\widetilde{{\swb}}_{S,i-1} \|^2 + \nn \\
& & \left( \frac{4}{1-\rho(T_{RR})} b^2 + 2c^2\right) \mu_\mathrm{max}^2\nn
\eq
Noting that $\rho(T_{RR})<1$ is independent of $\mu_{\max}$, we have
\be 1 -
\rho(T_{RR})-2\mu_\mathrm{max}^2\sigma_{RR}^2v^2
\beta_{R}^2 = O(1)\ee
for sufficiently small $\mu_\mathrm{max}$. Letting $i\rightarrow \infty$
we conclude that
\begin{equation*}
	\limsup_{i\to\infty} \Ex \| \bar{\swb}_{R,i} \|^2 \leq d \lim_{i\to\infty}\Ex \|\widetilde{\swb}_{S,i-1} \|^2 + O(\mu_\mathrm{max}^2)
\end{equation*}
where the scalar $d$ is defined by:
\be
d\define   \displaystyle \frac{\left( \frac{2\sigma_{SR}^2}{1 - \rho(T_{RR})} +
	2\mu_{\max}^2\beta_{S}^2\sigma_{SR}^2\right)}{1 - \rho(T_{RR})-2\mu_\mathrm{max}^2\sigma_{RR}^2v^2\beta_{R}^2} = O(1)
\ee
Since we know that $\limsup_{i\to\infty} \Ex \|\widetilde{{\swb}}_{S,i-1} \|^2 = O(\mu_\mathrm{max})$, we find that
\be
\limsup_{i\to\infty}
\Ex \|\bar{{\swb}}_{R,i} \|^2 = O(\mu_\mathrm{max})
\ee
Then, combining equation (\ref{eq.wBarTilde}) with above result we conclude that
\be
\limsup_{i\to\infty}\Ex \|\widetilde{\swb}_{R,i} \|^2 = O(\mu_\mathrm{max})
\ee

\section{Mean-Error Stability of (\ref{label.eq49})}\label{app.b}
\noindent We start by taking the expectation of both sides of (\ref{label.eq49}):
\bq
\Ex\bar{\swb}_{R,i} &=&  \mathcal{J}_{\epsilon}\tran \Ex \left[\left(I - \mathcal{V}_{\epsilon}\tran\mathcal{M}_{R}\bm{\cal H}_{R,i-1}  \mathcal{V}_{\epsilon}^{-\mathsf{T}}  \right)  \bar{\swb}_{R,i-1}\right] + \nn\\
&&		\Ex[\bar{\mathcal{T}}\tran_{SR} \left(I -\mathcal{M}_{S} \bm{\cal H}_{S,i-1}  \right)   \widetilde{\swb}_{S,i-1}] - \nn\\
&&	\bar{\mathcal{T}}\tran_{SR}\mathcal{M}_{S} b_{S} - \bar{\mathcal{T}}\tran_{RR}\mathcal{M}_{R}  b_{R}  \label{eq.EQ95}
\eq
where the terms involving the gradient noise are canceled out. Next, we need the
error quantities $\bm{\widetilde{\cal H}}_{S,i-1}$ and $\bm{\widetilde{\cal H}}_{R,i-1}$,which are defined in (\ref{eq.longHerrS}) and (\ref{eq.longHerrR}).

Using condition (\ref{jakd.187312}) and proceeding similarly to the argument that established (9.280) in \cite{Sayed14}, we can verify that
\bq
\|\bm{\widetilde{\cal H}}_{S,i-1}\| &\leq& \kappa_{S}\|\widetilde{\swb}_{S,i-1} \|   \label{eq.EQ121}\\
\|\bm{\widetilde{\cal H}}_{R,i-1}\| &\leq& \kappa_{R} \|\widetilde{\swb}_{R,i-1} \|  \label{eq.EQ102}
\eq
for some nonnegative constants $\kappa_{S}$ and $\kappa_{R}$. Substituting (\ref{eq.longHerrS}) and (\ref{eq.longHerrR}) back into (\ref{eq.EQ95}), we get:
\bq
\Ex\bar{\swb}_{R,i} &=&  \mathcal{J}_{\epsilon}\tran \left(I - \mathcal{V}_{\epsilon}\tran\mathcal{M}_{R}\mathcal {H}_{R}  \mathcal{V}_{\epsilon}^{-\mathsf{T}}  \right) \Ex  \bar{\swb}_{R,i-1} + \nn\\
&&		\bar{\mathcal{T}}\tran_{SR} \left(I -\mathcal{M}_{S} \mathcal {H}_{S}  \right)   \Ex\widetilde{\swb}_{S,i-1} - \nn\\
&&	\bar{\mathcal{T}}\tran_{SR}\mathcal{M}_{S} b_{S} - \bar{\mathcal{T}}\tran_{RR}\mathcal{M}_{R}  b_{R} + \nn\\
&&  \mathcal{J}_{\epsilon}\tran \Ex \left(\mathcal{V}_{\epsilon}\tran\mathcal{M}_{R} \bm{\widetilde{\cal H}}_{R,i-1}  \mathcal{V}_{\epsilon}^{-\mathsf{T}} \bar{\swb}_{R,i-1}\right)  + \nn \\
&&  \bar{\mathcal{T}}\tran_{SR} \Ex [ \mathcal{M}_{S} \bm{\widetilde{\cal H}}_{S,i-1} \widetilde{\swb}_{S,i-1}] \label{eq.EQ101}
\eq

Returning to (\ref{eq.EQ101}), we can now bound the norms of its last two terms as follows:
\be
\left\| \mathcal{J}_{\epsilon}\tran \Ex \left(\mathcal{V}_{\epsilon}\tran\mathcal{M}_{R} \bm{\widetilde{\cal H}}_{R,i-1}  \mathcal{V}_{\epsilon}^{-\mathsf{T}} \bar{\swb}_{R,i-1}\right) \right\|  \hspace{3.4cm}\nn \vspace{-0.3cm}
\ee
\bq
&\stackrel{(a)}{\leq}& \|\mathcal{J}_{\epsilon}\tran \|  \| \mathcal{V}\tran_{\epsilon} \| \| \mathcal{M}_{R} \| \|\Ex \bm{\widetilde{\cal H}}_{R,i-1}  \mathcal{V}_{\epsilon}^{-\mathsf{T}} \bar{\swb}_{R,i-1}  \| \nn\\
&\stackrel{(b)}{\leq}& \rho(T_{RR})  v' \mu_\mathrm{max} \Ex \| \bm{\widetilde{\cal H}}_{R,i-1} \| \|  \mathcal{V}_{\epsilon}^{-\mathsf{T}} \|  \|  \bar{\swb}_{R,i-1}  \| \nn \\
&\stackrel{(c)}{\leq}& \rho(T_{RR})  vv' \mu_\mathrm{max} \kappa_R   \Ex \| \widetilde{\swb}_{R,i-1} \| \Ex \| \bar{\swb}_{R,i-1} \| \nn \\
&\stackrel{(d)}{\leq}& \rho(T_{RR})  v^2v' \mu_\mathrm{max} \kappa_R  \sqrt{ \Ex \| \bar{\swb}_{R,i-1} \|^2}\,\sqrt{ \Ex \| \bar{\swb}_{R,i-1} \|^2} \nn \\
&\stackrel{(e)}{\leq}& \rho(T_{RR})  v^2v' \mu_\mathrm{max} \kappa_R  \Ex \| \bar{\swb}_{R,i-1} \|^2
\label{stepb.adlk12lk}\eq
where step (a) is due to the sub-multiplicative property of norms, step (b) uses Jensen's inequality, and introduces \be v'=\| \mathcal{V}\tran_{\epsilon} \|\ee
In addition, combining $ \|\mathcal{J}_{\epsilon}\tran\| =\sqrt{\rho({\mathcal{J}_{\epsilon}\mathcal{J}_{\epsilon}\tran})}$ with (\ref{usiadk..1lk3l1k2}), we also can conclude that,
\be
\|\mathcal{J}_{\epsilon}\tran\| \;\leq\; \rho(T_{RR})
\ee
Step (c)
uses (\ref{eq.EQ102}), step (d) uses (\ref{eq.wBarTilde}), and step (e) uses
$(\Ex \|\x\|)^2\leq \Ex\|\x\|^2$ for any random variable $\x$.  Similarly, we get:
\be
\left\| \bar{\mathcal{T}}\tran_{SR} \Ex \left( \mathcal{M}_{S} \bm{\widetilde{\cal H}}_{S,i-1} \widetilde{\swb}_{S,i-1} \right) \right\|  \leq
\sigma_{SR} \mu_\mathrm{max} \kappa_S \Ex \| \widetilde{\swb}_{S,i-1} \| ^2
\ee
For compactness of notation, we introduce the scalars:
\bq
e &\define& \rho(T_{RR}) v v' \mu_\mathrm{max} \kappa_R  = O(\mu_\mathrm{max}) \label{eq.scalar_e}\\
f &\define& \sigma_{SR} \mu_\mathrm{max} \kappa_S  = O(\mu_\mathrm{max})  \label{eq.scalar_f}
\eq
and return to equation (\ref{eq.EQ101}) to find that:
\bq
\| \Ex\bar{\swb}_{R,i} \| &\leq& \| \mathcal{J}_{\epsilon}\tran \|  \| I - \mathcal{V}_{\epsilon}\tran\mathcal{M}_{R}  \mathcal{H}_{R}  \mathcal{V}_{\epsilon}^{-\mathsf{T}}\| \|\Ex \bar{\swb}_{R,i-1}\| +\nn\\
&&	\| \bar{\mathcal{T}}\tran_{SR}\| \|I -\mathcal{M}_{S} \mathcal{H}_{S}\| \| \Ex \widetilde{\swb}_{S,i-1} \| + \nn \\
&&	\| \bar{\mathcal{T}}\tran_{SR}\mathcal{M}_{S} b_{S} - \bar{\mathcal{T}}\tran_{RR}\mathcal{M}_{R}  b_{R}  \| + \nn\\
&& 	e\Ex \|\bar{\swb}_{R,i-1}\|^2 + f \Ex \|\widetilde{\swb}_{S,i-1} \|^2  \nn \\&&\nn\\
&\leq&
\| \mathcal{J}_{\epsilon}\tran \|  \| I - \mathcal{V}_{\epsilon}\tran\mathcal{M}_{R}  \mathcal{H}_{R}  \mathcal{V}_{\epsilon}^{-\mathsf{T}}\| \|\Ex \bar{\swb}_{R,i-1}\| +\nn\\
&&	\| \bar{\mathcal{T}}\tran_{SR}\| \|I -\mathcal{M}_{S} \mathcal{H}_{S}\| \| \Ex \widetilde{\swb}_{S,i-1} \| + \nn \\
&& 	e\Ex \|\bar{\swb}_{R,i-1}\|^2 + f \Ex \|\widetilde{\swb}_{S,i-1} \|^2  +\nn \\
&& \mu_\mathrm{max} \left(\|\bar{\mathcal{T}}\tran_{SR}\|\|b_{S}\| + \|\bar{\mathcal{T}}\tran_{RR}\| \| b_{R}  \| \right)  \nn \\&&\nn\\
&\stackrel{(a)}{\leq}&
\rho(T_{RR}) \|\Ex \bar{\swb}_{R,i-1}\| + \sigma_{SR} \| \Ex \widetilde{\swb}_{S,i-1} \| +\nn \\
&& e\Ex \|\bar{\swb}_{R,i-1}\|^2 + f \Ex \|\widetilde{\swb}_{S,i-1} \|^2  +\nn \\
&& \mu_\mathrm{max} \left(\|\bar{\mathcal{T}}\tran_{SR}\|\|b_{S}\| + \|\bar{\mathcal{T}}\tran_{RR}\| \| b_{R}  \| \right)
\eq
In step (a), we used following inequality
\bq \| I -\mathcal{V}_{\epsilon}\tran\mathcal{M}_{R}
\mathcal {H}_{R} \mathcal{V}_{\epsilon}^{-\mathsf{T}}\| &\leq& 1 \label{eq.EQ74a}\\
\| I -\mathcal{M}_{S} \mathcal{H}_{S}\|&\leq &1 \label{eq.EQ75a}
\eq for sufficiently small  $\mu_{\max}$
and denoted the scalar coefficient here:
\be
b' \define \left(\|\bar{\mathcal{T}}\tran_{SR}\|\|b_{S}\| + \|\bar{\mathcal{T}}\tran_{RR}\| \| b_{R}  \| \right)  = O(1)
\ee
If we now let $i \to \infty$:
\bq
\limsup_{i \to \infty} \| \Ex\bar{\swb}_{R,i} \|  &\leq& \frac{\sigma_{SR}}{1-\rho(T_{RR}) } \| \Ex \widetilde{\swb}_{S,i-1} \| +  \nn \\
&&\frac{e}{1-\rho(T_{RR})} \Ex \|\bar{\swb}_{R,i-1}\|^2 + \nn\\
&&\frac{f}{1-\rho(T_{RR})}  \Ex \|\widetilde{\swb}_{S,i-1} \|^2  +  \nn\\
&&\frac{b'}{1-\rho(T_{RR})} \mu_\mathrm{max}
\eq
Notice that $\rho(T_{RR}) < 1$ and calling upon (\ref{kakd13}),  we conclude that
\bq
\limsup_{i \to \infty} \| \Ex\bar{\swb}_{R,i} \| = O(\mu_{\max})  \label{eq.yjwdx1}
\eq
Lastly, similar to equation equation (\ref{eq.wBarTilde}), we get following inequality,
\be
\|\Ex\widetilde{\swb}_{R,i}\| = \| \mathcal{V}_{\epsilon}^{-\mathsf{T}} \Ex \bar{\swb}_{R,i} \| \leq \| \mathcal{V}_{\epsilon}^{-\mathsf{T}} \|  \|\Ex \bar{\swb}_{R,i} \|  \label{eq.yjwdx2}
\ee
Combining (\ref{eq.yjwdx1}) and (\ref{eq.yjwdx2}) we arrive at
\be
\limsup_{i\to\infty}\|\Ex \widetilde{\swb}_{R,i}\|  = O(\mu_\mathrm{max}) \label{eq.yjwdx3}
\ee

\section{Stability of Fourth-Order Error Moment  of (\ref{label.eq49})}\label{app.c}
\noindent From recursion (\ref{label.eq49}) we obtain:
\be
\nn \Ex \|\bar{\swb}_{R,i} \|^4 \quad \quad \quad \quad \quad \quad \quad \quad \quad \quad \quad \quad \quad \quad \quad \quad \quad \quad \quad\quad
\ee
\bq	
&=&
\Ex \left\| \mathcal{J}_{\epsilon}\tran \left(I -\mathcal{V}_{\epsilon}\tran\mathcal{M}_{R} \bm{\cal H}_{R,i-1}
\mathcal{V}_{\epsilon}^{-\mathsf{T}} \right)  \bar{\swb}_{R,i-1}\right. + \nn \\
& & \;\;\;\;\; \bar{\mathcal{T}}\tran_{SR}\left( I -\mathcal{M}_{S} \bm{\cal H}_{S,i-1}  \right)   \widetilde{\swb}_{S,i-1} - \nn\\
& &\;\;\;\;\; \left. \bar{\mathcal{T}}\tran_{SR}\mathcal{M}_{S} b_{S} - \bar{\mathcal{T}}\tran_{RR}\mathcal{M}_{R}  b_{R}  \right\|^4 + \nn\\
& &  \Ex \left\|\bar{\mathcal{T}}\tran_{SR} \mathcal{M}_{S} \ssb_{S,i} +\bar{\mathcal{T}}\tran_{RR}\mathcal{M}_{R} \ssb_{R,i} \right\|^4 + \nn \\
& & 6 \; \Ex \left\| \mathcal{J}_{\epsilon}\tran \left(I -\mathcal{V}_{\epsilon}\tran\mathcal{M}_{R} \bm{\cal H}_{R,i-1}
\mathcal{V}_{\epsilon}^{-\mathsf{T}} \right)  \bar{\swb}_{R,i-1}\right. + \nn \\
& & \;\;\;\;\;\; \bar{\mathcal{T}}\tran_{SR}\left( I -\mathcal{M}_{S} \bm{\cal H}_{S,i-1}  \right)   \widetilde{\swb}_{S,i-1} -  \bar{\mathcal{T}}\tran_{SR}\mathcal{M}_{S} b_{S}\nn\\
& & \;\;\;\;\;\; \left.  - \bar{\mathcal{T}}\tran_{RR}\mathcal{M}_{R}  b_{R}  \right\|^2
\left\|\bar{\mathcal{T}}\tran_{SR} \mathcal{M}_{S} \ssb_{S,i} +\bar{\mathcal{T}}\tran_{RR}\mathcal{M}_{R} \ssb_{R,i} \right\|^2 \nn \\ \label{eq.ExpandForuth}
\eq
where we used the expansion
\[
\|a+b\|^4 = \|a\|^4+ \|b\|^4+6\|a\|^2\|b\|^2 + 4b\tran a\|a\|^2 + 4 a\tran b\|b\|^2
\]
and where the expectation of the last two terms in the expansion are eliminated since the gradient noise is zero mean and independent of other random variables.  Let us consider the last term in (\ref{eq.ExpandForuth}):
\begin{small}
\be
\mathrm{last\ term} \nn \hspace{7.5cm} \vspace{-0.2cm}
\ee
\begin{align}
\stackrel{(\ref{sunc.lakdq})}{\leq}&\;  6 \; \Big\{ \frac{\rho(\mathcal{J_{\epsilon}J_{\epsilon}}\tran)}{1-t} \Ex\| \bar{\swb}_{R,i-1}\|^2 + \frac{2\| \bar{T}_{SR} \|^2}{t} \Ex \| \widetilde{\swb}_{S,i-1} \|^2 +\nn \\
 &\; \;\;\;\frac{4}{t}\mu_\mathrm{max}^2\left( \|\bar{T}_{SR}\|^2 \|b_{S}\|^2 + \|\bar{T}_{RR}\|^2\|b_{R}\|^2  \right) \Big\}\nn\\
 &\; \times \; 2\; \mu_\mathrm{max}^2 \Ex \left\{ \|\bar{T}_{RR}\|^2 \| \ssb_{R,i}  \|^2 + \|\bar{T}_{SR}\|^2 \| \ssb_{S,i}  \|^2 \right\}  \nn \\ &\nn\\
\stackrel{(a)}{\leq}&\;
6 \; \Big\{ \frac{\rho(T_{RR})^2}{1-t} \Ex\| \bar{\swb}_{R,i-1}\|^2 + \frac{2\sigma_{SR}^2}{t} \Ex \| \widetilde{\swb}_{S,i-1} \|^2 +\nn \\
 &\; \;\;\;\frac{4}{t}\mu_\mathrm{max}^2 b^2  \Big\} \times \; 2\; \mu_\mathrm{max}^2  \left\{ \sigma_{SR}^2 \beta_{S}^2\, \Ex \| \widetilde{\swb}_{S,i-1}\|^2  + \right.\nn\\
 &\;\left. \sigma_{RR}^2 \beta_{R}^2\, \Ex \| \widetilde{\swb}_{R,i-1}\|^2  \|^2 + \|\bar{T}_{RR}\|^2 \sigma_{R}^2 + \|T_{SR}\|^2 \sigma_{S}^2 \right\}  \nn \\ &\nn\\
\stackrel{(b)}{\leq}&\;
6\Big\{\rho(T_{RR})\Ex \| \bar{\swb}_{R,i-1}\|^2  +  \nn \\ \nn
 &\;  \left( \frac{2\sigma_{SR} }{1-\rho(T_{RR}) } \right) \Ex \| \widetilde{\swb}_{S,i-1}\|^2  +\frac{4}{1-\rho(T_{RR}) }\mu_\mathrm{max}^2 b^2 \Big\} \times \\
 &\; 2\mu_\mathrm{max}^2 \left\{ \sigma_{SR}^2 \beta_{S}^2\, \Ex \| \widetilde{\swb}_{S,i-1}\|^2  + \sigma_{RR}^2 \beta_{R}^2\, \Ex \| \widetilde{\swb}_{R,i-1}\|^2 +c^2\right\} \label{eq.1a} 
\end{align}
\end{small}
In step (a), we expand the stochastic gradient noise similar to (\ref{eq.StoGradS})--(\ref{eq.StoGradR}),  and in step (b) we choose $t = 1 - \rho(T_{RR})$ and use the scalar defined (\ref{eq.ScalarC}).
To emphasize the order of the coefficients, we rewrite the above equation as:
\begin{align}
\mathrm{last\ term} \leq &\; \Big\{ O(1)\Ex \| \bar{\swb}_{R,i-1}\|^2 + O(1) \Ex \| \widetilde{\swb}_{S,i-1}\|^2  \nn \\
 &\;\;\; {}+O(\mu_\mathrm{max}^2 ) \Big\}\times \Big \{O(\mu_\mathrm{max}^2 )\Ex \| \bar{\swb}_{R,i-1}\|^2 +  \nn \\
 &\;\;\; O(\mu_\mathrm{max}^2) \Ex \| \widetilde{\swb}_{S,i-1}\|^2 + O(\mu_\mathrm{max}^2)\Big\}  \nn \\
= &\;  O(\mu_\mathrm{max}^2) \Ex \| \bar{\swb}_{R,i-1}\|^4 + O(\mu_\mathrm{max}^2)  \Ex \| \widetilde{\swb}_{S,i-1}\|^4  \nn \\
 &\;  \;\; {}+O(\mu_\mathrm{max}^4)  \label{eq.EQ113}
\end{align}
We consider next the second term using Jensen's inequality:
\bq
\mathrm{second\ term} &\leq & 8 \sigma_{SR}^4 \mu_\mathrm{max}^4 \| \ssb_{S,i} \|^4 + 8 \sigma_{RR}^4 \mu_\mathrm{max}^4 \| \ssb_{R,i} \|^4 \nn
\eq
Referring to equation (8.122) in \cite{Sayed14}, we also  conclude that
\bq
\Ex  \| \ssb_{S,i} \|^4 &\leq& \beta_{4,S}^2\, \Ex \| \widetilde{\swb}_{S,i-1}\|^4  + \sigma_{4,S}^2 \label{eq.4StoGradS}\\
\Ex  \| \ssb_{R,i} \|^4 &\leq& \beta_{4,R}^2\, \Ex \| \widetilde{\swb}_{R,i-1}\|^4  + \sigma_{4,R}^2 \label{eq.4StoGradR}
\eq
where
\bq
\beta_{4,S}^2 &=&\max_{1\leq k\leq N_{gS}}\,\beta_{4,k}^2 \;=\;O(1)\\
\beta_{4,R}^2 &=&\max_{N_{gS}+1\leq k\leq N}\,\beta_{4,k}^2 \;=\;O(1)\\
\sigma_{4,S}^2 &=& \sum_{k=1}^{N_{gS}} \sigma^2_{4,k} \;=\;O(1) \\
\sigma_{4,R}^2 &= &\sum_{k=N_{gS}+1}^{N} \sigma^2_{4,k} \;=\;O(1)
\eq
Hence, we find for the second term with simplified notation that,
\begin{align}
\mathrm{second\ term}  \leq &\; O(\mu_\mathrm{max}^4) \Ex \| \widetilde{\swb}_{S,i-1}\|^4 + \nn\\
&\; O(\mu_\mathrm{max}^4) \Ex \| \bar{\swb}_{R,i-1}\|^4 + O(\mu_\mathrm{max}^4) \label{eq.EQ117}
\end{align}
Lastly, we consider the first term using Jensen's inequality again:
\bq
\mathrm{first\ term} &\leq & \frac{ \rho(T_{RR})^4}{(1-t)^3} \Ex \| \bar{\swb}_{R,i-1}\|^4 +  \nn \\
& &\frac{ 8\sigma_{RR}^4}{t^3}  \Ex \| \widetilde{\swb}_{S,i-1}\|^4 + \nn \\
& & \frac{64}{t^3}\mu_\mathrm{max}^4\left( \sigma_{RR}^4\|b_{R}\|^4 + \sigma_{SR}^4 \|b_{S}\|^4 \right)
\eq
for any $0<t<1$. Here, we select $t=1-\rho(T_{RR})$ so that
\bq
\mathrm{first\ term} &\leq & \left(\rho(T_{RR})\right)\Ex \| \bar{\swb}_{R,i-1}\|^4 +   \\
& &\frac{ 8\sigma_{RR}^4}{(1-\rho(T_{RR}))^3}  \Ex \| \widetilde{\swb}_{S,i-1}\|^4 + \nn \\
& & \hspace{-0.4cm}\frac{64}{(1-\rho(T_{RR}))^3}\mu_\mathrm{max}^4\left( \sigma_{RR}^4\|b_{R}\|^4 + \sigma_{SR}^4 \|b_{S}\|^4 \right)  \nn
\eq
Recalling that $\rho(T_{RR}) < 1$, we get:
\bq
\mathrm{first\ term} &\leq & \rho(T_{RR}) \Ex \| \bar{\swb}_{R,i-1}\|^4 + O(1) \Ex \| \widetilde{\swb}_{S,i-1}\|^4 + \nn\\
&&O(\mu_\mathrm{max}^4)\label{eq.EQ119}
\eq
Substituting (\ref{eq.EQ113}), (\ref{eq.EQ117}), and (\ref{eq.EQ119}) back into (\ref{eq.ExpandForuth}),  we find:
\bq
\Ex \| \bar{\swb}_{R,i}\|^4 &=& \Big(\rho(T_{RR}) + O(\mu_\mathrm{max}^2)\Big)\Ex \| \bar{\swb}_{R,i-1}\|^4 + \nn\\
& & \;\;O(1)\Ex \| \widetilde{\swb}_{S,i-1}\|^4 +O(\mu_\mathrm{max}^4)
\eq
Recalling $\rho(T_{RR}) < 1$ again, we then let $i \to \infty$ to find:
\be
\limsup_{i\to\infty} \Ex \| \bar{\swb}_{R,i}\|^4 = O(1)\limsup_{i\to\infty} \Ex \| \widetilde{\swb}_{S,i-1}\|^4 +O(\mu_\mathrm{max}^4) \nn
\ee
Since we already know that $ \limsup_{i\to\infty} \Ex \| \widetilde{\swb}_{S,i-1}\| = O(\mu_\mathrm{max}^2)$, we conclude that
\be
\limsup_{i\to\infty}\Ex \| \bar{\swb}_{R,i}\|^4  = O(\mu_\mathrm{max}^2)
\ee
Combining the square of equation (\ref{eq.wBarTilde}) with the above result we arrive at
\be
\limsup_{i\to\infty}\Ex \| \widetilde{\swb}_{R,i}\|^4  = O(\mu_\mathrm{max}^2)
\ee

\section{Proof of Theorem~\ref{theorem.longterm}}\label{app.longterm}
To simplify the notation, we introduce the differences
\bq
\z_{S,i}  &\define& \widetilde{\swb}'_{S,i} - \widetilde{\swb}_{S,i} \\
\z_{R,i}  &\define& \widetilde{\swb}'_{R,i} - \widetilde{\swb}_{R,i} 
\eq
Subtracting recursions (\ref{eq:ErrRec}) and (\ref{eq:ErrRec2}), we get:
\bq
\ba{c}\z_{S,i} \\ \z_{R,i}\ea = {\cal A}\tran
\left(
I
-
\ba{cc}
\mathcal{M}_{S} {\cal H}_{S} & \\
& \mathcal{M}_{R} {\cal H}_{R}
\ea
\right) 
\ba{c}\z_{S,i} \\ \z_{R,i}\ea\nn\\ {}+ {\cal A}\tran {\cal M}\ba{c} \bm{\cal \widetilde{H}}_{S,i} \widetilde{\swb}_{S,i}  \\ \bm{\cal \widetilde{H}}_{R,i} \widetilde{\swb}_{R,i}   \ea
\eq
so that
\bq
\z_{R,i}\hspace{-0.1cm}  &=&\hspace{-0.1cm}  \mathcal{T}\tran_{RR}\left(I -\mathcal{M}_{R} {\cal H}_{R}  \right)  	 \z_{R,i-1} + \nn\\
&&\hspace{-0.1cm} 		\mathcal{T}\tran_{SR} \left(I -\mathcal{M}_{S}  {\cal H}_{S}  \right)   	 \z_{S,i-1} + \nn\\
&& \hspace{-0.1cm} \mathcal{T}\tran_{RR} \mathcal{M}_{R} \bm{\cal \widetilde{H}}_{R,i} \widetilde{\swb}_{R,i}  +  \mathcal{T}\tran_{SR} \mathcal{M}_{S}\bm{\cal \widetilde{H}}_{S,i} \widetilde{\swb}_{S,i} 
\eq
Using the eigen-decomposition (\ref{eq.63}), we can transform the above equation into.
\bq
\bar{\z}_{R,i}\hspace{-0.1cm}  &=&\hspace{-0.1cm}   \mathcal{J}\tran_{\epsilon}\left(I -{\cal V}_{\epsilon}\tran\mathcal{M}_{R} {\cal H}_{R}{\cal V}^{-\mathsf{T}}  \right)  	\bar \z_{R,i-1} + \nn\\
&&\hspace{-0.1cm} 		\mathcal{\bar{T}}\tran_{SR} \left(I -\mathcal{M}_{S}  {\cal H}_{S}  \right)   	 \z_{S,i-1} + \nn\\
&&\hspace{-0.1cm}  \mathcal{\bar{T}}\tran_{RR} \mathcal{M}_{R}  \bm{\cal \widetilde{H}}_{R,i} \widetilde{\swb}_{R,i}  +  \mathcal{\bar{T}}\tran_{SR} \mathcal{M}_{S}\bm{\cal \widetilde{H}}_{S,i} \widetilde{\swb}_{S,i} 
\eq
where we are using the notation from (\ref{eq.def67})  as well as  
\be
\bar{\z}_{R,i} \define {\cal V}_{\epsilon}\tran \z_{R,i} 
\ee
Now note that
\bq
\|\mathcal{\bar{T}}\tran_{SR} \mathcal{M}_{S} \bm{\cal \widetilde{H}}_{S,i} \widetilde{\swb}_{S,i}  \|^2 &\leq& \|\mathcal{\bar{T}}\tran_{SR} \|^2 \| \mathcal{M}_{S} \|^2 \| \bm{\cal \widetilde{H}}_{S,i} \widetilde{\swb}_{S,i}  \|^2 \nn\\
&\stackrel{(\ref{eq.scalar1})}{=}& \mu_{\max}^2 \sigma_{SR}^2 \| \bm{\cal \widetilde{H}}_{S,i} \widetilde{\swb}_{S,i}   \|^2  \nn\\
&\stackrel{(\ref{eq.EQ121})}{\leq}& \mu_{\max}^2 \sigma_{SR}^2\kappa_{S}^2 \|\widetilde{\swb}_{S,i} \|^4
\eq
Similarly, we can verify that
\bq
\|\mathcal{\bar{T}}\tran_{RR} \mathcal{M}_{R}  \bm{\cal \widetilde{H}}_{R,i} \widetilde{\swb}_{R,i}  \|^2 &\leq& \mu_{\max}^2 \sigma_{RR}^2\kappa_{R}^2 \|\widetilde{\swb}_{R,i} \|^4
\eq
Therefore, following steps similar to the arguments in Appendices \ref{app.a}  and \ref{app.b} we then conclude that 
\be
\limsup_{i\to\infty} \Ex\| \z_{R,i} \|^2 = O(\mu_{\max}^2)
\ee
which is equivalent to:
\be
\limsup_{i\to\infty} \Ex\| \widetilde{\swb}'_{R,i} - \widetilde{\swb}_{R,i} \|^2 = O(\mu_{\max}^2)
\ee
Finally, note that 
\bq
\Ex\| \widetilde{\swb}'_{R,i}\|^2 &=& \Ex\| \widetilde{\swb}'_{R,i} - \widetilde{\swb}_{R,i} + \widetilde{\swb}_{R,i}\|^2 \nn \\
&\leq&  \Ex\| \widetilde{\swb}'_{R,i}- \widetilde{\swb}_{R,i}\|^2 + \Ex\|\widetilde{\swb}_{R,i}\|^2 +\nn\\
& & 2|\Ex (\widetilde{\swb}'_{R,i}-\widetilde{\swb}_{R,i})^* \widetilde{\swb}_{R,i}|
\eq
Hence, we conclude that
\be
\limsup_{i\to\infty} \Ex\| \widetilde{\swb}_{R,i}\|^2 = \limsup_{i\to\infty} \Ex\| \widetilde{\swb}'_{R,i}\|^2  + O(\mu_{\max}^{3/2})
\ee

\section{Proof of Theorem~\ref{theorem.3}}\label{app.d}
We know from Theorem 11.2 in \cite{Sayed14} that the MSD expression for a generic agent
$k$ in a connected network is given by:
\be
\mbox{\rm MSD}_k= \Tr  ( \mathcal{E}_{k} \mathcal{X} )  \label{eq.genMSD}
\ee
where $\mathcal{E}_{k}$ is a block diagonal matrix with zero blocks except for an
identity matrix of size $M \times M$ at the location corresponding to agent $k$. Moreover, the matrix ${\cal X}$ is defined as:
\bq
\mathcal{X} &=& \sum_{n=0}^{\infty} \mathcal{B}^n\mathcal{Y}{\mathcal{B}\tran}^{n} \\
\mathcal{Y} &=& \mathcal{A}\tran \mathcal{MSMA} \label{eq.defY}\\
{\cal S}&=& \mbox{\rm diag}\{G_1,G_2,\ldots,G_N\}\\
{\cal B}&=&{\cal A}\tran(I-{\cal M}{\cal H})\\
{\cal M}&=&\mbox{\rm diag}\{{\cal M}_S,{\cal M}_R\}\\
{\cal H}&=&\mbox{\rm diag}\{{\cal H}_S,{\cal H}_R\}\\
{\cal H}_S&=&\mbox{\rm diag}\{H_1,H_2,\ldots,H_{N_{gS}}\}\\
{\cal H}_R&=&\mbox{\rm diag}\{H_{N_{gS}+1},\ldots,H_{N}\}
\eq
Using the block Kronecker product notation, it can be verified that expression (\ref{eq.genMSD})
can be equivalently expressed in the form:
\be
\mbox{\rm MSD}_k =  \left( \mathrm{bvec} (\mathcal{Y}\tran) \right)\tran (I-\mathcal{F})^{-1}  \mathrm{bvec} (\mathcal{E}_{k})  \label{eq.genMSD2}
\ee
where \be \mathcal{F} \define \mathcal{B}\tran \otimes_b \mathcal{B}\tran\ee
and $\mbox{\rm bvec}(Z)$ denotes the block vectorization operation; for a matrix $Z$ with blocks of size
$M\times M$, this operation vectorizes each $M\times M$ submatrix of $Z$ and the resulting
vectors are subsequently stacked on top of each other. We now examine how expression (\ref{eq.genMSD}) simplifies for both cases of agents in group $S$ and agents in
group $R$. To do so, we introduce the canonical Jordan decompensation of matrix $A$. Recall that $A$ has
$S$ leading primitive left-stochastic matrices, $\{A_{s},\,s=1,2,\ldots,S\}$. Each of these matrices has all its
eigenvalues strictly inside the unit circle with the exception of a single eigenvalue equal to one. The
corresponding Perron vector is denoted by $p_s$ and has dimensions $N_s\times 1$. We therefore conclude that
the canonical Jordan decomposition of
$A$ has the form:
\be
A=V\cdot \ba{ccc}I_{S}&\vline&\\\hline & \vline&J_{\epsilon}'\ea V^{-1}
\label{kadlk173812a}\ee
All eigenvalues of
$J'_{\epsilon}$ are strictly smaller than one.  We partition $V$ in the form
\be
V\define \ba{ccc}P&\vline&V_{x}\ea
\ee
where $P$ is $N\times S$ and $V_{x}$ collects the remaining right-eigenvectors of $A$. We also partition
$V^{-1}$ as
\be
V^{-1}\define \ba{c}Q\tran\\\hline V_{y}\ea
\ee
where the matrices $\{Q\tran,\,V_{y}^{-1}\}$ contain the corresponding
left-eigenvectors and $Q\tran$ is $S\times N$. We can use expression (\ref{label.eq30}) for the limiting power of
$A$ to identify the matrices $\{P,Q\}$ in the above decomposition.  Indeed, if
we compute the limit of $A^n$ using  (\ref{kadlk173812a}) we conclude that
\bq
\lim_{n\rightarrow\infty}\,A^n&=&\lim_{n\rightarrow\infty}\,\left(V\cdot \ba{cc}I_{S}&\\& {J_{\epsilon}'}^{n}
\ea\cdot V^{-1}\right)\nn\\
&=&PQ\tran
\eq
Comparing with (\ref{label.eq30}), we conclude that it must hold
\be
\ba{ccc}\Theta&\vline&\Theta W\\\hline 0&\vline&0\ea\;=\;PQ\tran
\ee
so that we can set
\bq
P&=&\ba{cccc}p_1\\&p_2\\&&\ddots\\&&&p_S\\ \hline\\[1pt]
\multicolumn{4}{c} {\raisebox{1ex}[0pt]{\Large 0}}
\ea \label{eq.defP}\\
Q\tran&=&\ba{cc}L\tran&L\tran W\ea \label{eq.defQ}
\eq
where we are defining
\be L\tran \define \mbox{\rm blockdiag}\left\{\one_{N_1}^{\sf T},\one_{N_2}^{\sf T}, \ldots,\one_{N_S}^{\sf T}
\right\}
\ee
We conclude that
\bq
\mathcal{A} &=& {\cal V}\ba{c|c}
I_{S M} &  \\ \hline
& \mathcal{J'_{\epsilon}}
\ea {\cal V}^{-1}\nn\\
&=&
\ba{c|c}
\mathcal{P} & \mathcal{V}_{x}
\ea
\ba{c|c}
I_{S M} &  \\ \hline
& \mathcal{J'_{\epsilon}}
\ea
\ba{c}
\mathcal{Q}\tran \\ \hline
\mathcal{V}_{y}	\ea\label{eq.JCFA}
\eq
where
\bq
{\cal V}&=& V\otimes I_M\\
{\cal P}&=&P\otimes I_M\\
{\cal Q}\tran&=&Q\tran\otimes I_M\\
{\cal J}'_{\epsilon}&=&J_{\epsilon}'\otimes I_M\\
{\cal V}_{x}&=&V_x\otimes I_M\\
{\cal V}_{y}&=&V_y\otimes I_M
\eq
Using the Jordan decomposition (\ref{eq.JCFA}) for ${\cal A}$ we can now write: {\small
	\bq
	\mathcal{B}\tran&=& (I-{\cal M}{\cal H}){\cal A}\nn\\
	&=&{\cal A}-{\cal M}{\cal H}{\cal A}\nn\\
	&=&{\cal V}\left\{\ba{cc}I_{SM}&\\&{\cal J}'_{\epsilon}\ea-{\cal V}^{-1}{\cal M}{\cal H}{\cal V}
	\ba{cc}I_{SM}&\\&{\cal J}'_{\epsilon}\ea\right\}{\cal V}^{-1}\nn\\
	&=&{\cal V}\ba{ccc}I_{SM}-{\cal D}_{11}&\vline&
	-{\cal D}_{12}\\\hline -{\cal D}_{21} & \vline& {\cal J}'_{\epsilon}-
	{\cal D}_{22}\ea {\cal V}^{-1}   \label{eq.expandB}
	\eq
}
\noindent where
\bq
{\cal D}_{11}&=&{\cal Q}\tran{\cal M}{\cal H}{\cal P}\nn\\
&=&\mbox{\rm blockdiag}\left\{\sum_{k=1}^{N_1} q_{1,k} H_{1,k},  \ldots, \sum_{k=1}^{N_S} q_{S,k} H_{S,k} \right\} \nn\\
&=&O(\mu_{\max})\\\nn\\
{\cal D}_{12}&=&{\cal Q}\tran{\cal M}{\cal H}{\cal V}_x{\cal J}'_{\epsilon}\;=\;O(\mu_{\max})\\
{\cal D}_{21}&=&{\cal V}_y {\cal M}{\cal H}{\cal P}\;\;\;=\;O(\mu_{\max})\\
{\cal D}_{22}&=&{\cal V}_y{\cal M}{\cal H}{\cal V}_x{\cal J}'_{\epsilon}\;=\;O(\mu_{\max})
\eq
We then obtain that
\be
\mathcal{F} = (\mathcal{V} \otimes_b \mathcal{V } )\mathcal{Z}
(\mathcal{V } \otimes_b \mathcal{V } )^{-1}
\ee
where we introduced the matrix
\begin{small}
\be
\mathcal{Z} \define
\ba{cc}I_{SM}-{\cal D}_{11}&
-{\cal D}_{12}\\ -{\cal D}_{21} & {\cal J}_{\epsilon}-
{\cal D}_{22}\ea
\otimes_b
\ba{cc}I_{SM}-{\cal D}_{11}&
-{\cal D}_{12}\\ -{\cal D}_{21} & {\cal J}_{\epsilon}-
{\cal D}_{22}\ea
\ee
\end{small}
This relation allows us to determine a low-rank expansion for $(I-{\cal F})$ as follows. First note that
\be
(I-\mathcal{F})^{-1} = (\mathcal{V } \otimes_b \mathcal{V} )(I-\mathcal{Z})^{-1}
(\mathcal{V} \otimes_b \mathcal{V} )^{-1} \label{eq.78}
\ee
Then, we can appeal to the derivation used in the proof of Lemma 9.5 from \cite{Sayed14}
to conclude that the entries of $(I-{\cal Z})^{-1}$ are in the order of:
\be
(I-\mathcal{Z})^{-1} =
\ba{c|c}
O(1/{\mu_\mathrm{max}}) & O(1) \\ \hline
O(1) & O(1)
\ea
\ee
where the size of the leading block is $(SM)^2 \times (S M)^2$. Since $O(1/\mu_\mathrm{max})$
dominates $O(1)$ for sufficiently small $\mu_\mathrm{max}$,
we have that
\be
(I-\mathcal{Z})^{-1} =
\ba{c|c}
\left( I \otimes {\cal D}_{11} + {\cal D}\tran_{11}\otimes I \right) ^{-1} & 0\\ \hline
0 & 0
\ea + O(1) \label{eq.80}
\ee
Noting that the Hessian matrices $\{H_{s,k}\}$ are symmetric, we conclude that ${\cal D}\tran_{11}={\cal D}_{11}$. Substituting (\ref{eq.80}) into (\ref{eq.78}), we arrive at the following low-rank approximation:
\begin{align}
	&(I-\mathcal{F})^{-1} \label{eq.lowf} \\
	& = \left( \mathcal{P} \otimes_b \mathcal{P} \right) \left( I \otimes {\cal D}_{11}
	+ {\cal D}_{11}\otimes I \right) ^{-1}
	\left( \mathcal{Q}\tran \otimes_b \mathcal{Q}\tran \right) + O(1) \nn 
\end{align}
If we now substitute (\ref{eq.lowf}) into (\ref{eq.genMSD2}), we obtain:
\begin{align}
	\left( \mathrm{bvec} (\mathcal{Y}\tran) \right)\tran (I-\mathcal{F})^{-1}  \mathrm{bvec} (\mathcal{E}_{k}) = O(\mu_{\max}^2) + \nn \quad \quad \quad \quad \quad \quad \\
	\left( \mathrm{bvec} (\mathcal{Y}\tran) \right)\tran \left( \mathcal{P} \otimes_b
	\mathcal{P} \right)  {\cal Z}_{1}^{-1}
	\left( \mathcal{Q}\tran \otimes_b \mathcal{Q}\tran \right)\mathrm{bvec} (\mathcal{E}_{k})  \label{eq.MSDcomplicated}
\end{align}
where \be {\cal Z}_{1} \define  I \otimes {\cal D}_{11} + {\cal D}_{11}\otimes I\ee
We can simplify the right-hand side of (\ref{eq.MSDcomplicated}) as follows.  Starting from the
rightmost term we note that:
\be
\left( \mathcal{Q}\tran \otimes_b \mathcal{Q}\tran \right) \mathrm{bvec} (\mathcal{E}_{k}) = \mathrm{bvec}\left(\mathcal{Q}\tran \mathcal{E}_{k} \mathcal{Q}\right) \hspace{1.8cm}
\ee \vspace{-0.6cm}
\bq
= \mathrm{bvec}\left( \left[Q\tran {E}_k Q\right] \otimes I_M\right) \quad\quad\quad\quad \quad\quad\quad\quad \quad\quad\quad\quad \quad\quad\quad\nn\\
\stackrel{(a)}{=}
\begin{cases}
	\mathrm{bvec} (E_s \otimes I_M)\hspace{-0.1cm} &(\text{\mbox{\rm when $k\in$ sub-network $s$ in group $S$}}) \\
	\mathrm{bvec} (c_{k} c\tran_{k} \otimes I_M)\hspace{-0.1cm} & (\text{\mbox{\rm when $k\in$ sub-network $r$ in group $R$}})
\end{cases}\nn
\eq
where the matrix $E_{k}$ is an $S\times S$ matrix with all zero entries except at location
$(k,k)$, where the entry is equal to one. Moreover, the   vector $c_{k}$ is the same defined earlier in (\ref{eq.definef}).
In step (a), we expanded the quantity $Q\tran E_k Q$ using the following identity:
\bq
Q\tran E_k Q &=& \sum_{i=1}^{N}\sum_{j=1}^{N} E_k(i,j) [Q\tran]_{:,i} [Q]_{j,:} \\
&\stackrel{(b)}{=}& [Q\tran]_{:,k} [Q]_{k,:}  \nn \\
&\stackrel{(\ref{eq.defQ})}{=}&
\begin{cases}
	[L\tran]_{:,k}[L]_{k,:} &(\text{\mbox{\rm when $k\in$  group $S$}}) \\
	[L\tran W]_{:,k}[W\tran L ]_{k,:} & (\text{\mbox{\rm when $k\in$ group $R$}})
\end{cases}\nn
\eq
where the notation $[\ \cdot \ ]_{:,i}$ and  $[\ \cdot \ ]_{j,:}$ denotes here the the $i-$th column and $j-$th row of the matrix, respectively. Step (b) is because of the definition of $E_s$

It is therefore clear that the performance of agents in group $S$ will be different from the
performance  of agents in group $R$. For agents in group $S$, from this point onwards,
we can follow the same argument from  Lemma 11.3 in \cite{Sayed14} to arrive at their MSD expression.  With regards to agents in group $R$ we proceed as follows. Let
\be x_1 = {\cal Z}_{1}^{-1} \mathrm{bvec} (c_{k} c\tran_{k} \otimes I_M))\ee
or, equivalently, by using the definition of ${\cal Z}_{1}$:
\be
(I \otimes {\cal D}_{11})x_1 + ( {\cal D}_{11} \otimes I )x_1 =
\mathrm{bvec} \left(c_{k} c\tran_{k} \otimes I_M\right)
\ee
Let ${\cal X}_{1} = \mathrm{unbvec}(x_1)$ denote the $SM \times SM $ matrix whose block
vector representation is $x_1$. Then, the matrix ${\cal X}_1$ is the solution to the Lyapunov equation:
\be
{\cal D}_{11}{\cal X}_{1}+{\cal X}_{1}{\cal D}_{11} = c_{k} c\tran_{k} \otimes I_M
\ee
We can solve for the block diagonal entries of ${\cal X}_1$. For example, consider the $s-$th diagonal block of size $M\times M$. It satisfies:
\be
\left(\sum_{k=1}^{N_{s}} q_{s,k} H_{s,k}\right)  \left[{\cal X}_{1}\right]_{s,s} +  \left[{\cal X}_{1}\right]_{s,s} \left(\sum_{k=1}^{N_{s}} q_{s,k} H_{s,k}\right)
= c_{k}^2(s) I_M
\ee
so that
\be
\left[{\cal X}_{1}\right]_{s,s} = \frac{1}{2} c_{k}^2(s) \left( \sum_{k=1}^{N_{s}} q_{s,k} H_{s,k} \right) ^{-1}
\ee
\noindent Since this information is sufficient to obtain the MSD level for agent $k$, we do not need to compute the off-diagonal blocks of ${\cal X}_1$.

Returning to equation (\ref{eq.MSDcomplicated}), and using properties of the block Kronecker product operation, we have
\be
\left( \mathrm{bvec} (\mathcal{Y}\tran) \right)\tran [\mathcal{P} \otimes_b \mathcal{P} ] \mathrm{bvec}({\cal X}_{1}) \quad\quad\quad\quad\quad\quad\quad\quad \quad \quad\quad\quad\quad\quad\nn
\ee
\bq
&=& \Tr  \left[ \mathrm{unbvec} \{ ( \mathcal{P} \otimes_b \mathcal{P} )\  \mathrm{bvec}({\cal X}_{1}) \} \mathcal{Y} \right] \nn \\
&= &\Tr  \left[  \mathcal{P}  {\cal X}_{1} \mathcal{P}\tran \mathcal{Y} \right] \nn \\
&\stackrel{\tiny (\ref{eq.defY})}{=}& \Tr  \left[  {\cal X}_{1}  \mathcal{P}\tran( \mathcal{A}\tran \mathcal{MSMA} )\mathcal{P} \right] \nn\\
&\stackrel{\rm (a)}{=}& \Tr  \left[  {\cal X}_{1} \mathcal{U}\tran \mathcal{SU} \right] \nn\\
&\stackrel{\rm (b)}{=}& \frac{1}{2} \Tr   \left[\sum_{s=1}^{S} c_{k}^2(s)\left(
\sum_{k=1}^{N_{s}}q_{s,k}H_{s,k} \right)^{-1}\left( \sum_{k=1}^{N_{s}}q_{s,k}^2{G}_{s,k}\right)\right] \nn\\
\eq
where in step (a) we introduced
\be
\mathcal{U} \define \mathcal{M AP} \stackrel{(\ref{definqa.s}) }{=}
\ba{cccc}
q_1 & & & \\
&q_2 & & \\
& & \ddots&\\
& & & q_S \\ [1pt]
\hline\\[1pt]
\multicolumn{4}{c} {\raisebox{1ex}[0pt]{\Large 0}} \ea
\otimes I_M
\ee
In step (b), we exploited the fact that $\mathcal{U}\tran \mathcal{SU}$
is block diagonal matrix.

\bibliographystyle{IEEEbib}
\bibliography{refs}

\end{document}

%% file: macros.tex
\newcommand{\sw}{{\scriptstyle{\mathcal{W}}}}
\newcommand{\swb}{{\scriptstyle{\boldsymbol{\mathcal{W}}}}}
\newcommand{\ssb}{{\scriptstyle{\boldsymbol{\mathcal{S}}}}}

\newtheorem{lemma}{{Lemma}}
\newtheorem{theorem}{{Theorem}}

\def\tran{^{\mathsf{T}}}
\def\one{\mathds{1}}

\newcommand{\bp}{\small \begin{proof}}
\newcommand{\ep}{\end{proof} \normalsize}

\newcommand{\Ex}{\mathbb{E}\hspace{0.05cm}}
\newcommand{\bm}[1]{\mbox{\boldmath $#1$}}

\newcommand{\be}{\begin{equation}}
\newcommand{\ee}{\end{equation}}
\newcommand{\bq}{\begin{eqnarray}}
\newcommand{\eq}{\end{eqnarray}}
\newcommand{\bqn}{\begin{eqnarray*}}
\newcommand{\eqn}{\end{eqnarray*}}
\newcommand{\nn}{\nonumber}
\newcommand{\ba}{\left[ \begin{array}}
\newcommand{\ea}{\\ \end{array} \right]}

\newcommand{\define}{\;\stackrel{\Delta}{=}\;}
\newcommand{\Tr}{\mbox{\rm {\small Tr}}}

\def\H{{\boldsymbol{H}}}

\def\d{{\boldsymbol{d}}}

\def\h{{\boldsymbol{h}}}

\def\s{{\boldsymbol{s}}}

\def\u{{\boldsymbol{u}}}
\def\v{{\boldsymbol{v}}}
\def\w{{\boldsymbol{w}}}
\def\x{{\boldsymbol{x}}}
\def\y{{\boldsymbol{y}}}
\def\z{{\boldsymbol{z}}}

\def\real{{\mathbb{R}}}

\def\Zint{{\mathchoice{\setbox1=\hbox{\sf Z}\copy1\kern-.75\wd1\box1}
{\setbox1=\hbox{\sf Z}\copy1\kern-.75\wd1\box1}
{\setbox1=\hbox{\scriptsize\sf Z}\copy1\kern-.75\wd1\box1}
{\setbox1=\hbox{\scriptsize\sf Z}\copy1\kern-.75\wd1\box1}}}

\makeatletter
\def\hlinewd#1{%
  \noalign{\ifnum0=`}\fi\hrule \@height #1 \futurelet
   \reserved@a\@xhline}
\makeatother